\documentclass[12pt,preprint]{aastex}

\newcommand{\hbetaa}{H$\beta$}
\newcommand{\mgfepp}{[MgFe]$^\prime$}
\newcommand{\afee}{[$\alpha$/Fe]}

\newcommand{\mgtwoo}{Mg$_2$}

\newcommand{\hbeta}{H$\beta$~}
\newcommand{\mgfep}{[MgFe]$^\prime$~}
\newcommand{\afe}{[$\alpha$/Fe]~}
\newcommand{\feavg}{$<$Fe$>$~}
\newcommand{\mgtwo}{Mg$_2$~}

\newcommand{\etal}{et al.~}

\shorttitle{The Globular Cluster System of NGC 4636}
\shortauthors{Park et al.}

\begin{document}
\title{The Globular Cluster System of NGC 4636
and Formation of Globular Clusters in Giant Elliptical Galaxies\altaffilmark{*}} 

\author{ Hong Soo Park\altaffilmark{1}, Myung Gyoon Lee\altaffilmark{1},
Ho Seong Hwang\altaffilmark{2}, 
Sang Chul Kim\altaffilmark{3},
Nobuo Arimoto\altaffilmark{4}, Yoshihiko Yamada\altaffilmark{4},
Naoyuki Tamura\altaffilmark{5,6}, and Masato Onodera\altaffilmark{7} 
}

\email{hspark@astro.snu.ac.kr, mglee@astro.snu.ac.kr,
hhwang@cfa.harvard.edu, sckim@kasi.re.kr, 
arimoto.n@nao.ac.jp, yoshihiko.yamada@nao.ac.jp,
naoyuki.tamura@ipmu.jp, and monodera@phys.ethz.ch
}

\altaffiltext{1}{Astronomy Program, Department of Physics and Astronomy, Seoul National University, Korea}
\altaffiltext{2}{Smithsonian Astrophysical Observatory, 60 Garden Street,
Cambridge, MA 02138, USA}
\altaffiltext{3}{Korea Astronomy and Space Science Institute, Daejeon 305-348, Korea}
\altaffiltext{4}{National Astronomical Observatory of Japan, Tokyo, Japan}
\altaffiltext{5}{Kavli Institute for Physics and Mathematics of the Universe, University of Tokyo, Kashiwa-city 277-8583, Japan}
\altaffiltext{6}{Subaru Telescope, National Astronomical Observatory of Japan, Hilo, USA}
\altaffiltext{7}{Institute	for	Astronomy,	ETH	Z\"urich,	Wolfgang-Pauli-strasse	27,	8093	Z\"urich,	Switzerland}
\altaffiltext{*}{Based on data collected at Subaru Telescope, which is operated by the National Astronomical Observatory of Japan.}

\begin{abstract}

We present a spectroscopic analysis of
  the metallicities, ages, and alpha-elements of the globular clusters (GCs)
  in the giant elliptical galaxy (gE) NGC 4636 in the Virgo cluster.
Line indices of the GCs  are
  measured from the integrated spectra obtained with
  Faint Object Camera and Spectrograph (FOCAS) on the Subaru 8.2 m Telescope.
We derive [Fe/H] values of 59 GCs 
  based on the Brodie \& Huchra method, and
  [Z/H], age, and \afe values of 33 GCs 
  from the comparison of the Lick line indices with
   single stellar population models.
The metallicity distribution of NGC 4636 GCs shows a hint of a bimodality
   with two peaks at 
   $\textrm{[Fe/H]}=-1.23 (\sigma=0.32)$ and $-0.35 (\sigma=0.19)$. 
The age spread is large from 2 Gyr to 15 Gyr and
  the fraction of young GCs with age $<$ 5 Gyr is about 27\%.
The \afe of the GCs shows a broad distribution with a mean value \afe $\approx 0.14$ dex.
The dependence of these chemical properties on the galactocentric radius is weak.
We also derive the metallicities, ages, and \afe values for the GCs 
  in other nearby gEs (M87, M49, M60, NGC 5128, NGC 1399, and NGC 1407)
  from the line index data in the literature  using the same methods
  as used for NGC 4636 GCs.
The metallicity distribution of GCs in the combined sample of seven gEs including NGC 4636 
  is found to be bimodal, 
  supported by the KMM test with a significance level of $>$99.9\%.
All these gEs harbor some young GCs with ages less than 5 Gyr.
  The mean age of the metal-rich GCs ([Fe/H] $> -0.9$) is about 3 Gyr younger 
  than that of the metal-poor GCs. 
The mean value of \afe of the gE GCs is 
   smaller than that of the Milky Way GCs.
We discuss these results 
  in the context of GC formation in gEs.
\end{abstract} 


\section{Introduction}

It is generally believed that the stars in a globular cluster (GC) are born almost 
  at the same time and have similar metal abundances, 
  even though there are some GCs with multiple stellar populations in the Milky Way (MW)
  \citep{mari09,lee09,bel10}.
GCs are often assumed as single stellar population (SSP) systems
  and their observed quantities can be compared with theoretical models 
  for the estimation of their physical parameters.
They also contain information on the formation and evolution of their host galaxies.
Therefore, analysis of their ages and metallicities
  can provide critical clues for understanding the evolution of their host galaxies.

According to the current models of structure formation,
  giant elliptical galaxies (gEs) are considered to form and evolve
  through mergers with various galaxies.
They might have accreted thousands of GCs and sometimes formed new star clusters during the merging process \citep{lee03, bro06, lee10a}. 
Therefore, investigation of ages and metallicities of the GCs in gEs 
  may allow us to understand how GCs and their host galaxies formed and evolved.
Though, in the case of gEs, it is not possible to obtain detailed information 
  on the resolved stars in GCs with current telescopes, 
  we can effectively estimate their age and metallicity using integrated spectroscopy.

There are several studies on the ages and metallicities of GCs in nearby gEs based on integrated spectroscopy, 
  which are summarized briefly below for six gEs (NGC 5128, M87, M49, M60, NGC 1407, NGC 1399). Basic properties of these galaxies are given in Table 3 of \citet{lee10a}.
   
The GCs of the nearest gE NGC 5128 were extensively studied 
  thanks to its relatively short distance, 3.8 Mpc \citep{pen04, bea08, woo10, woo10b}.
Recently, \citet{woo10} derived the metallicities, ages and \afe  values for 72 GCs  in NGC 5128 using the Lick indices obtained with the Gemini/GMOS. %
Ninety-two percent 
  of the metal-poor GCs (MP GCs) have ages older than 8 Gyr, while 56 percent of the metal-rich GCs (MR GCs) do so.
They found that the metallicity distribution of the NGC 5128 GCs  is bimodal   if seen through the metallicities derived directly from the spectroscopic Lick indices ([MgFe]$^\prime$),  
  but the distribution is not well constrained if they use Lick indices and SSP models.
From these results together with the [$\alpha$/Fe] data, 
  they  suggested  
  that most GCs in NGC 5128 formed rapidly at early times, 
  while some GCs formed later with subsequent major accretion.

The GCs of M87 (NGC 4486), a cD galaxy located at the center of the Virgo cluster,
  were studied spectroscopically early by \citet{coh98}.
They reported that the mean metallicity of the GCs is 
  higher than that of the MW GCs,
  and that the metallicity distribution shows a bimodal feature with peaks at
  [Fe/H] = --1.3 and --0.7.
The metallicities of the M87 GCs show a small radial gradient with a large scatter, 
  but the ages of the GCs show no radial gradient.
They suggested that the median age of the M87 GCs is 13 Gyr, similar to that of the MW GCs.

M49 (NGC 4472) is the brightest gE in the Virgo cluster. 
\citet{coh03} presented the spectroscopic results of 47 GCs in this galaxy obtained with the Keck telescope.
They concluded that the M49 GCs have slightly higher metallicities than the M87 GCs,
  while other metallicity and age parameters are basically similar to those of the M87 GCs.
They estimated that the mean age of these GCs is about 10 Gyr.

\citet{pie06} reported a study of 38 GCs in M60 (NGC 4649) in the Virgo cluster.
They found that several young (2 -- 3 Gyr old) GCs in this galaxy have supersolar metallicities,
  while the majority of the GCs have a wide range of metallicity ([Fe/H] = --2.0 to 0.0),
  and are older than 10 Gyr.  
They also found that 
  the alpha-element ratios of the M60 GCs decrease with increasing metallicity.

The GCs in NGC 1407, the brightest galaxy in the NGC 1407 subgroup of the Eridanus group, were studied recently by \citet{cen07}.
They found that 20 GCs are mostly old, 
  while three GCs could be young ($\sim 4$ Gyr).
The metallicity of the GCs 
  reaches up to slightly above solar and
  the mean \afe ratio is $\sim 0.3$ dex, similar to that of the MW GCs.

Spectroscopic properties of the GCs in NGC 1399, 
  gE 
  in the center of the Fornax cluster,
  were studied early by \citet{kis98} and \citet{for01}. 
From the Keck spectra of 10 GCs,
  \citet{for01} showed that at least some of the GCs 
  have supersolar metal abundances.
Two GCs with significantly higher \hbeta values and enhanced abundance ratios
  could be regarded as young ($\sim 2$ Gyr) or extremely old ($\sim 15$ Gyr) ages
  with a warm blue horizontal branch.
From this
  they proposed a possibility of a complicated age distribution among the MR GCs
  and of some constraints on their chemical enrichment at later epochs.  

These previous studies have some limitations.
  The number of the studied galaxies in this regard is only six, and
  the number of GCs in each galaxy is still small.  
  The areal coverage for some galaxies is not wide enough, either.
Therefore, it is needed to obtain new spectroscopic data 
 for more GCs in each galaxy and for more gEs
in order to better understand the chemical properties of gE GCs.

We carried out a spectroscopic study of GCs in another Virgo gE NGC 4636.
NGC 4636, located at the southeast boundary of the Virgo, 
	harbors GC populations with a wide spatial distribution \citep{lee10b}. 
	It is one of the best targets 
	for the study of age and metal abundance 
	using the integrated spectroscopy together with M87, M49, and M60.
There are several previous studies on 
  the photometry, kinematics, and X-ray imaging of NGC 4636 GCs (summarized below),
 but there 
 is no  published study of the metallicities, 
  ages, and alpha-elements
  for the GCs based on spectroscopy.

The specific frequency of NGC 4636 GCs is $S_{N} = 5.6 - 8.9$, 
  larger than those of other elliptical galaxies in the Virgo Cluster \citep{kis94,dir05}.
 The GCs of NGC 4636 show a bimodal color distribution,   while brighter GCs have intermediate colors
 \citep{dir05}. The number of blue GCs is larger than that of red GCs.
  The number of the red GCs connected with low-mass X-ray binaries  is about two times 
  larger than that of the blue GCs \citep{kim06,pos09}.

Using the radial velocity data 
  for NGC 4636 GCs, 
\citet{sch06} reported that the dark matter fraction within one effective radius 
   is 20\% $\sim$ 30\%. However, 
 \citet{cha08} suggested that the dark matter distribution is 
  highly concentrated toward the inner halo.
 \citet{lee10a} also confirmed the need of a large amount of dark matter.
 \citet{sch06} and \citet{lee10a} found that the velocity dispersion for the blue GCs 
    is slightly larger than that for the red GCs.
 They  also studied the radial variation, the rotation, and the orbit 
    of NGC 4636 GC sub-populations.

We have been carrying out a project to investigate the spectroscopic
  properties of the GCs in nearby galaxies
  to understand the formation of GCs in galaxies. 
Our study on the kinematics of the GC system of M60 was
  presented in \citet{lee08b} and \citet{hwa08}, and
  another on the GC system of the spiral galaxy M31 was given in 
  \citet{kim07} and \citet{lee08c}.
Recently, we presented the measurement of radial velocities for
  the GCs in NGC 4636 in a companion paper \citep{par10}, 
  and detailed kinematic analysis of these data in \citet{lee10a}.
  We also presented a spectroscopic study of the M86 GCs \citep{par12b}.
  Here we present the chemical analysis of the NGC 4636 GCs.
  By combining 
  the GC data of seven gEs
  (including NGC 4636) available in the literature, 
we also investigate the chemical properties of the gE GCs.

This paper is composed as follows.
Section 2 describes briefly the spectroscopic observation and data reduction.
In \S3 
  we explain how we determine metallicities, ages, and alpha-elements of the GCs
   using the line indices measured in the optical spectra. 
\S 4 presents the chemical analysis results of NGC 4636 GCs. 
In \S 5, we compare the chemical properties of NGC 4636 GCs
   with those of other gEs,
  and discuss the implication of the results. 
Primary results are summarized in \S 6.

\section{Observation and Data Reduction}\label{data}

We carried out a spectroscopic observation of GCs in NGC 4636, using
  the Faint Object Camera and Spectrograph (FOCAS; \citealt{kas02})
  at the Subaru 8.2 m Telescope.
The spectra were obtained in Multi-Object Spectrography (MOS) mode of FOCAS.
Target selection, spectroscopic observation, data reduction, and velocity measurement
  were described in detail in \citet{par10}.
We obtained flux-calibrated spectra of the targets 
  using BD+33d2642, a standard star for flux calibration taken together with the targets. 
For the calibration of line indices, 
  we used the spectra of five MW GCs    (M5, M13, M92, M107, and NGC 6624)
  observed in the long slit mode during the same observing run. 

\begin{figure}
\plotone{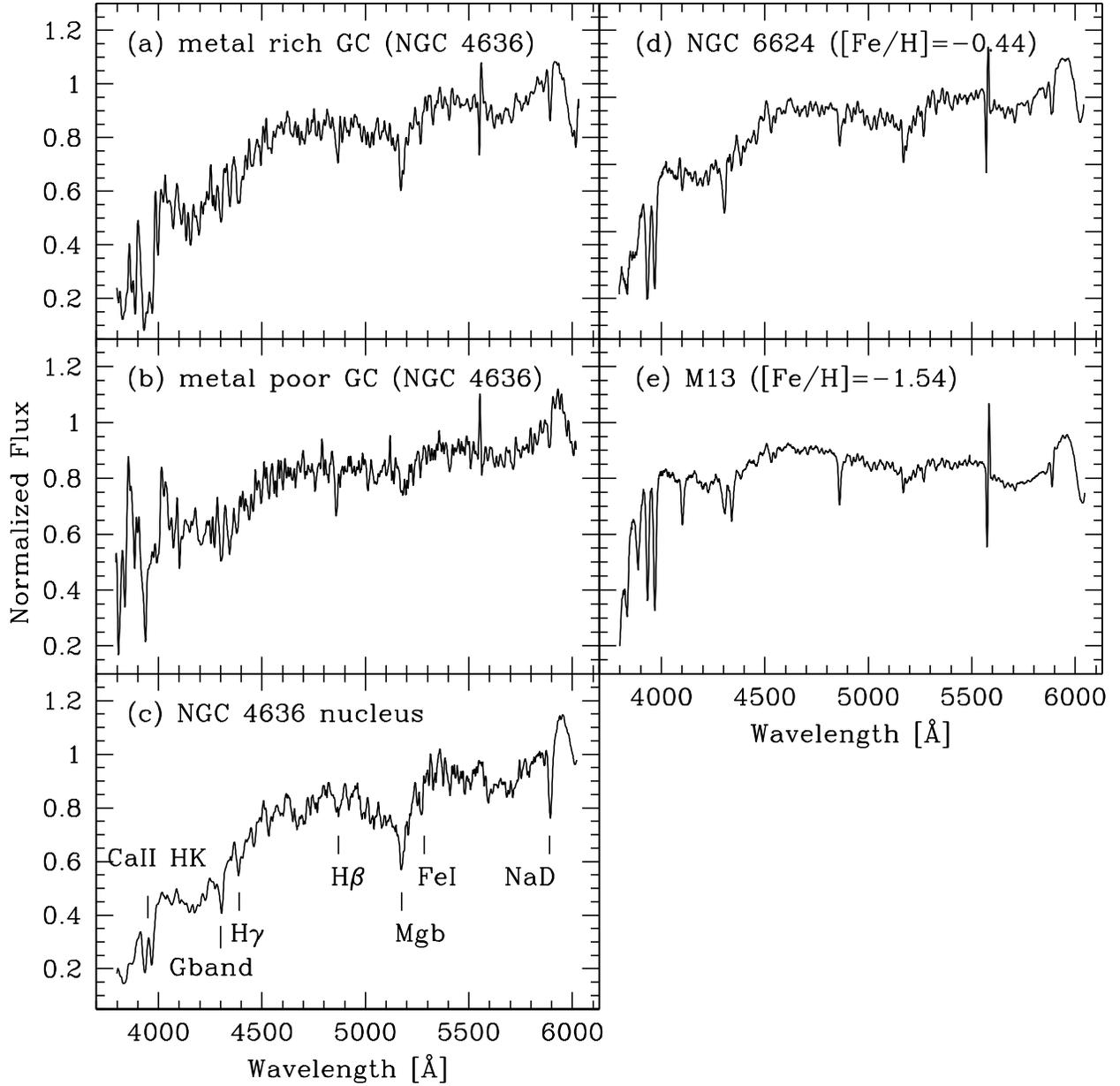}
\caption{
Example spectra of
(a) a metal-rich (MR) GC in NGC 4636 (ID: 154)  with $T_1=19.82$ and [Fe/H] = $-$0.39 dex;
(b) a metal-poor (MP) GC in NGC 4636 (ID: 9995) with $T_1=19.22$ and [Fe/H] = $-$1.49 dex;
(c) NGC 4636 nucleus;
(d) NGC 6624, a MR MW GC with [Fe/H] = $-$0.44 dex; and
(e) M13, a MP MW GC with [Fe/H] = $-$1.54 dex.
All spectra are plotted in the rest frame and 
are smoothed with the Lick resolution.
\label{fig-spec}}
\end{figure}

In Figure \ref{fig-spec}, we show the flux-calibrated spectra of 
an MR GC and an MP GC in NGC 4636, NGC 6624 and M13 in the MW Galaxy,
and the NGC 4636 nucleus.
These spectra are normalized at about 5400 \AA.
These spectra show strong absorption lines typically seen in GCs 
  such as H$\beta$, Mgb, and FeI.
MR GCs (Figure \ref{fig-spec} (a) and (d)) have strong absorption lines at Mgb,
while MP GCs (Figure \ref{fig-spec} (b) and (e)) have relatively weak lines.
Note that there are some sky-line residuals at $\sim 5577$ \AA,
  which are less than 2\% of their original sky-lines.
  Although these residuals remain because of 
  incomplete distortion correction in the optics, they affect little the index analysis in this study.   

\section{Metallicity, Age, and [$\alpha$/Fe] Measurement}\label{analysis}

\subsection{Line index measurement}

We measured two kinds of absorption line indices from the spectra of NGC 4636 GCs: 
  (1) the \citet{bro90} line indices, used for determining the metallicities of GCs
  with empirical relations between absorption line indices and metallicities,
  and (2) the Lick line indices, used for measuring 
  the metallicities, ages, and alpha-elements from the comparison
  with SSP models.
Following subsections describe the measurement of each index.
 
\subsubsection{Brodie \& Huchra (BH) line indices}

\citet{bro90} and \citet{huc96} presented linear relations 
  between absorption line strength indices obtained from the integrated spectra of old stellar systems
  and their mean metallicity to determine the metal abundance of the systems. 
Their method 
  was developed to minimize the systematic effects such as 
  reddening, individual element abundance anomalies, and instrumental effects.
They used the spectra and infrared colors of the MW GCs, M31 GCs, and NGC 188 giant stars 
  for the metallicity calibration of the observed spectra. 

We measured the absorption line indices from the flux-calibrated spectra of NGC 4636 GCs
  after shifting each spectrum to the rest frame
  following the prescription of \citet{bro90} and \citet{huc96}.
The measured absorption line indices are calibrated to the BH index system
  with a zero point offset, $Index({\rm BH})=Index({\rm Subaru})+ constant$, 
  determined from the spectra of five MW GCs common in this study and \citet{huc96}.
The offsets (the $constant$ in the above equation) we derived are
  $0.014 \pm 0.015$ for G band, 
  $-0.013 \pm 0.009$ for MgH, 
  $-0.021 \pm 0.018$ for Mg2, 
  and $0.008 \pm 0.009$ for Fe5270. 
Among the 105 NGC 4636 GCs with measured velocities \citep{par10}, 
  we derived the BH line indices for 59 GCs with S/N $\gtrsim 10$ and $T_1<21$.
The resulting index errors are smaller than 
  0.10 mag for G band, 0.03 mag for MgH and Mg2, and 0.04 mag for Fe5270.

\subsubsection{Lick line indices}

Lick absorption line indices are 
  useful for measuring the metallicity and age of old stellar systems
  from the integrated spectra with low/medium resolution 
  \citep{puz02, puz05, bea08, tra08, woo10}. 
We measured the Lick line indices as follows.
We first smoothed our spectra with the Lick resolution \citep{wor97}
  after shifting each spectrum to the rest frame.
Then we derived the Lick line indices from the 
  spectra of NGC 4636 GCs,
  following the definitions and the methods 
  given
  in \citet{wor94} and \citet{wor97}.
Line index errors are derived from the photon noise in the spectra
  before the flux calibration.
Then we calibrated the resulting line indices to the Lick system with the zero point offset,
  $Index({\rm Lick})=Index({\rm Subaru})+constant$, 
  determined from the spectra of five MW GCs common in this study,
  \citet{tra98}, and \citet{kun02}.

Table \ref{tab-translick} lists 
  the zero point offsets derived in this study.
The first, second, third, and fourth column are  
  the name of the Lick index, unit of each index, zero point offset, and $rms$, respectively.
Note that these offsets are slightly different from those in \citet{pau10}; 
  these differences could systemically lead to 
  the uncertainties of 
  2.3 Gyr in age, 0.14 dex in metallicity, and 0.18 dex in alpha-elements. 
  We finally measure the Lick line indices for 33 GCs in NGC 4636, 
  listed in Tables \ref{tab-lickindex1} and \ref{tab-lickindex2}. 
  The associated errors are in Tables \ref{tab-lickerr1}  and  \ref{tab-lickerr2}.  
We use only the spectra with S/N $\gtrsim 15$,
  where the S/N value is measured at the continuum part of the Lick line indices   such as \hbetaa, \mgtwoo, Mgb, Fe5270, and Fe5335.

\subsection{Determination of Metallicity, Age, and [$\alpha$/Fe]}

We derived metallicity using three methods (the Brodie \& Huchra method, the Lick index grid method and the $\chi^2$ minimization method), and age and [$\alpha$/Fe] using two methods (the Lick index grid method and the $\chi^2$ minimization method) for NGC 4636 GCs. Each method is described below.

\subsubsection{Brodie \& Huchra (BH) method}

We use the method described in \citet{bro90} to derive metallicities 
  from the absorption line indices for NGC 4636 GCs. 
They 
  recommended six best indices (G band, MgH, Mg2, Fe5270, CNB, and $\Delta$) 
  as the primary calibrators among the 12 line indices  
  for the empirical relations between the BH line indices and the metallicities. 
We used, however, only four primary line indices to determine the metallicities 
  of NGC 4636 GCs, 
  excluding CNB and $\Delta$ because of low S/N values in their wavelength range.
The relations between metallicity and the four indices 
  are as follows.
  $\textrm{[Fe/H]}_{G band}=11.415\times \textrm{Gband}-2.455$,
  $\textrm{[Fe/H]}_{MgH}=20.578 \times \textrm{MgH}-1.840$,
  $\textrm{[Fe/H]}_{Mg2}=9.921 \times \textrm{Mg2}-2.212$, and
  $\textrm{[Fe/H]}_{Fe5270}=20.367 \times \textrm{Fe5270}-2.086$.
We then derive the final metallicity value of each GC by taking 
an error-weighted average of four measurements.
Here the error is the mean of the standard deviation.
The mean error for 59 GCs derived from this method is $0.31\pm0.16$ dex 
  with the maximum error of about 0.6 dex.

\subsubsection{Lick index grid method} 

To obtain the estimates of metallicity, age, and \afe 
  from the integrated spectra of the NGC 4636 GCs,
  we use the Lick index grid method (the grid method hereafter),
  which is to derive the age and abundance 
  from the comparison of the Lick indices with the line index grids 
  predicted from SSP models \citep{tri95, tra00, tho04, puz05}.
  Here, we adopt SSP models given by \citet{tho03, tho04, tho05}.

We use the grids of \mgfep versus \hbeta provided by \citet{tho03} 
  to estimate the metallicity and age of each GC.
  The composite index \mgfepp, 
  defined as
  \mgfep$= \sqrt{\rm{Mgb} \times\ (0.72 \times \rm{Fe5270} + 0.28 \times \rm{Fe5335})}$,
  is a good tracer of metallicity because of  little sensitivity to \afe \citep{tho03}.
  We use \hbeta as an age indicator, 
  which is the least sensitive to \afe among the Balmer lines \citep{tho03}.
We also use the grids of \mgtwo versus \feavg in order to derive  \afe estimates.
   The \mgtwo index is very sensitive to \afee, and
   the \feavg index, defined as \feavg $=$ (Fe5270+Fe5335)/2,
   is a metallicity indicator on this grid \citep{tho03}. 

\begin{figure}
\plotone{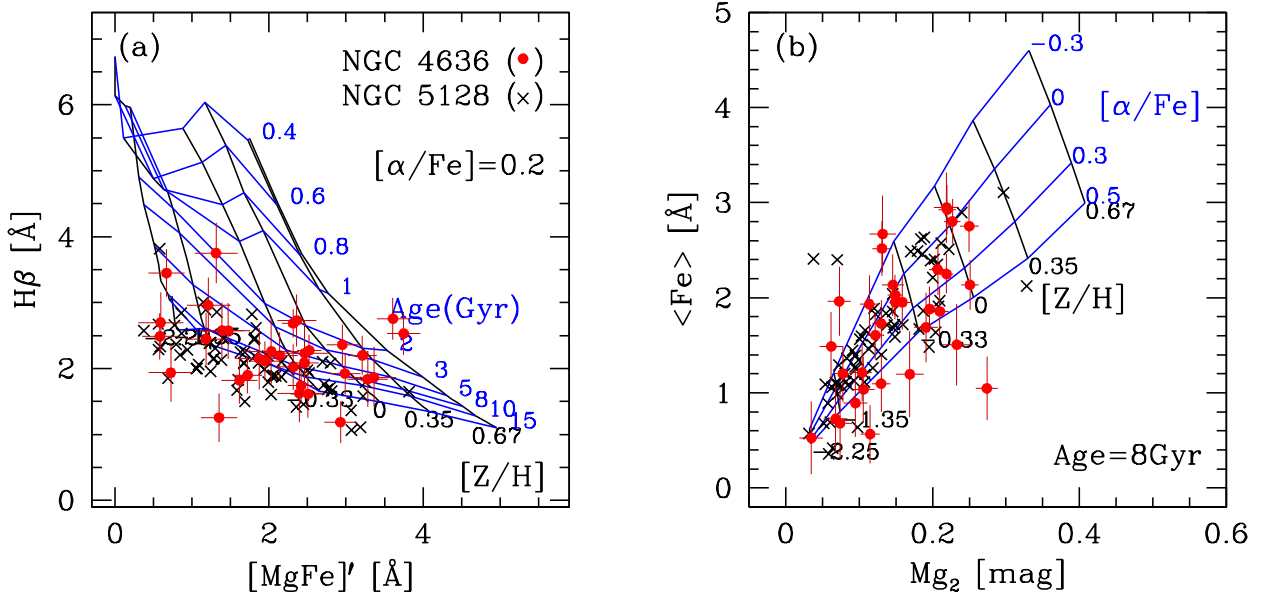}
\caption{Lick line index diagrams. 
(a) \hbeta line versus \mgfep for NGC 4636 GCs measured in this study (circles).
The crosses represents NGC 5128 GCs \citep{woo10}. 
The grids represent the SSP models 
  for various values of [Z/H] (--2.25, --1.35, --0.33, 0, 0.35, and 0.67) and ages (0.4, 0.6, 0.8, 1, 2, 3, 5, 8, 10, and 15 Gyr) given by \citet{tho03,tho04}. 
(b) $<$Fe$>$ versus Mg$_2$. The grids are for a model age of 8 Gyr.
\label{fig-samplegrid}}
\end{figure}

Figure \ref{fig-samplegrid} shows 
   the measured 
   indices in comparison with SSP model grids for various values of [Z/H] and age.
Panel (a) displays the Lick line indices 
  of \hbeta versus \mgfep for the NGC 4636 GCs 
  measured  in this study as well as 
  those for the NGC 5128 GCs \citep{woo10} 
  for comparison. 
Grids indicate several values of [Z/H] (--2.25, --1.35, --0.33, 0.0, 0.35, and 0.67 dex) and ages (0.4, 0.6, 0.8, 1, 2, 3, 5, 8, 10, and 15 Gyr) from the SSP model with  \afe = 0.2.
The NGC 4636 GCs show
  a large dispersion in ages; 
  most of them are old 
  and some GCs are younger than 5 Gyr.
The GCs in NGC 4636 and in NGC 5128 show similar age distributions,
  while the latter have a relatively small number of young GCs than the former. 
Metallicities of GCs in NGC 4636 and in NGC 5128
  show a wide dispersion at [Z/H] $<0.7$ dex.
%
%
Panel (b) displays \feavg versus Mg$_2$,
  with the SSP grids for age $= 8$ Gyr (mean age of our sample).
While the scatter looks large, 
  most GCs in NGC 4636 are concentrated on [$\alpha$/Fe] of $\sim 0$ and $\sim 0.5$.
GCs in NGC 5128 also show a large dispersion and
  most of them are located at [$\alpha$/Fe] $\sim 0$.
In the followings, quantitative analyses 
  on the age, metallicity, and \afe of NGC 4636 GCs
  are shown.

Since
  the dependence of the \hbeta versus \mgfep grid on \afe is not negligible,
  we need \afe information for accurate estimation of age and metallicity.
  An \mgtwo versus \feavg grid also needs an age value 
  for accurate estimation of \afe and metallicity.
Thus we need an iteration between the two grids 
  in order to determine age, metallicity, and \afe of an object
  more accurately.
We follow the iteration technique described in \citet{puz05}, which is summarized in
the following.
First, the age and metallicity of one GC is determined from the \hbeta versus \mgfep grid 
  in a given \afe as in Figure \ref{fig-samplegrid} (a).
Next, we determine the \afe and metallicity 
  using the \feavg versus \mgtwo grid with a previously determined age 
  as in Figure \ref{fig-samplegrid} (b).
Then, we derive again the age and metallicity of the GC 
  using the \hbeta versus \mgfep grid at the \afe value obtained above. 
We repeat this iteration until the [Z/H] difference converges to smaller than 0.01 dex.
Finally, 
the converged values are selected as final values of age, metallicity, and \afe of the GC.

To estimate the errors of the age, metallicity, and \afe values of a GC,
  we calculate ages, metallicities, and \afe values 
   of the four data points 
   composed of 
  \hbeta $\pm$ error and \mgfep $\pm$ error in the \hbeta versus \mgfep grid 
  and \feavg $\pm$ error and \mgtwo $\pm$ error in the \feavg versus \mgtwo grid.
The difference between the average of these four values and the estimate directly calculated
  from the $index$ is taken as the final error
  for the age, metallicity, and [$\alpha$/Fe] of the GC. 
For some GCs outside the model grid, we adopt the value of 
  the nearest envelope of the model grid 
  in the direction of error vector 
  as their estimates following \citet{puz05}. 
For example, if the data points are below the grid limit in the \hbeta versus \mgfep diagram, 
  we adopt the maximum value of age (i.e. 15 Gyr) at [Z/H] $>-0.5$ dex, 
  but the age value of about 12 Gyr due to some overlapped grids at [Z/H] $<-0.5$ dex. 
For the data points below the grid limit in the \feavg versus \mgtwo diagram, 
  we also adopt the maximum value of \afe (i.e. 0.5 dex).

\subsubsection{$\chi^2$ minimization method} 

\citet{pro04} suggested 
a method that can measure the physical parameters (metallicity, age, and alpha-elements) for GCs in a robust way
based on simultaneous fitting of all the available Lick line indices with $\chi^2$ minimization.
We call this method the $\chi^2$ minimization method hereafter. 
This method maximizes the use of available information, while the grid method uses only a pair of indices.
Recently this method was used in the studies of 
  the GCs in gE M60 by \citet{pie06} and
  early-type dwarf galaxies in Virgo by \citet{pau10}.

  We applied this method to the NGC 4636 GCs using the same SSP models as used in the grid method.
  Among 10 Lick line indices except Mg$_1$ at $4500-5750$ \AA, 
  we used about eight indices after $\sim 2 \sigma$ clipping of their $\chi$.  
  The error of each parameter is the value 
  where $\chi^2$ difference above the minimum reaches $1 \sigma$ significance level. 
  We list the measured metallicities, ages, and \afe for the NGC 4636 GCs in Table \ref{tab-catalog}.
  
%

\section{Results}\label{result}

\subsection{A Catalog of Metallicity, Age, and [$\alpha$/Fe]}
Among the 105 GCs 
  whose radial velocities are measured in \citet{par10},
we obtained [Z/H], age, and \afe for 33 GCs with the grid method and $\chi^2$ minimization method,
  while we obtained [Fe/H] for 59 GCs with the BH method.
Table \ref{tab-catalog} lists 
  metallicities, ages, and [$\alpha$/Fe] values for the NGC 4636 GCs.

\begin{figure}
\plotone{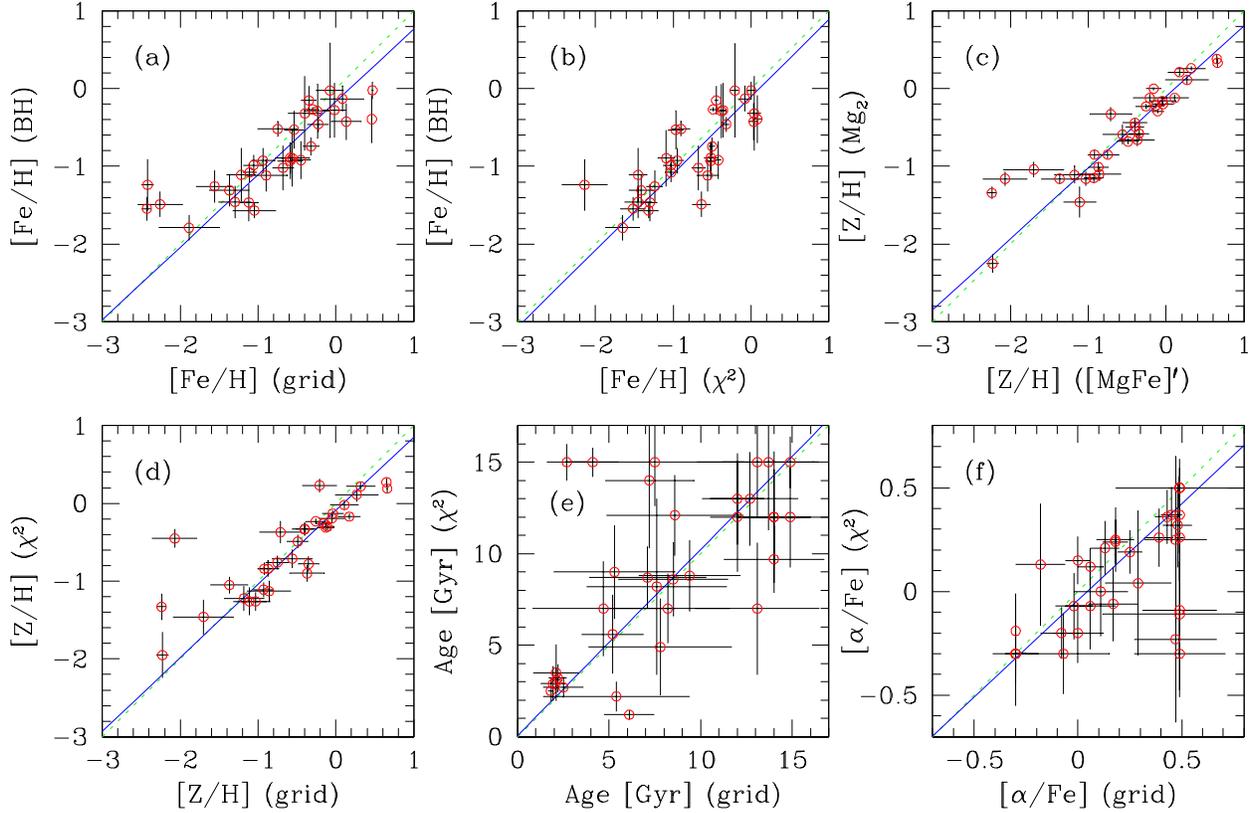}
\caption{ 
(a) Comparison of metallicities for NGC 4636 GCs measured with the BH method and
with the grid method. 
Here the grid metallcities are the values derived from the \hbeta versus \mgfep grid.
(b) Comparison of [Fe/H] for NGC 4636 GCs measured with 
the BH method and the $\chi^2$ minimization method 
(c) Comparison of [Z/H] values obtained from 
  \mgtwo grids and \mgfep grids.
Comparison of parameters for NGC 4636 GCs measured with 
the grid method and the $\chi^2$ minimization method: (d) [Z/H], (e) age, and (f) \afee. 
Here the grid metallcities are the values derived from the \hbeta versus \mgfep grid. 
The solid and dotted lines represent 
  the linear least-squares fit with 2 $\sigma$ clipping  and the one-to-one relation, respectively.  
\label{fig-compmetal}}
\end{figure}

Figure \ref{fig-compmetal} shows a comparison
of the results for NGC 4636 GCs derived from three methods.
Panel (a) shows a comparison
  of the metallicities obtained from the BH method and the grid method.
[Fe/H](grid or $\chi^2$) is converted from the total metallicity, [Z/H], 
  using the correlation between the two: 
  $\textrm{[Fe/H]}=\textrm{[Z/H]}-0.94$ \afe \citep{tho03}. 
  Here we adopt \afe = 0.2,  
  which is the mean \afe for NGC 4636 GCs.
Panel (a) shows a good agreement
  between the two measurements. 
  We derive a linear least-squares fit: 
  $\textrm{[Fe/H]}{\rm (BH)}=0.935(\pm0.080)\textrm{[Fe/H]}{\rm (grid)}+0.170(\pm0.069)$ with $rms=0.227$.
In the metal poor region, however, 
  the metallicities derived from the grid method for three GCs are smaller than 
  those from the BH method.
  %
Panel (b) illustrates a comparison
  of the metallicities obtained from the BH method and the $\chi^2$ minimization method,
  showing a good agreement
  between the two measurements. 
  We derive a linear least-squares fit:
  $\textrm{[Fe/H]}{\rm (BH)}=0.988(\pm0.072)\textrm{[Fe/H]}{\rm ( \chi^2 )}-0.105(\pm0.078)$ with $rms=0.297$.
%
  
Panel (c) displays a relation 
  between [Z/H] derived from the \hbeta versus \mgfep grid 
  and that from the \feavg versus \mgtwo grid.
It also shows a good agreement between the two estimates except for three outliers.
  We derive a linear least-squares fit:
  $\textrm{[Z/H]}{\rm (Mg}_2)=0.914(\pm0.058)\textrm{[Z/H]}{\rm ([MgFe]}^\prime)+0.105(\pm0.031)$ with $rms=0.192$. 
For the following analysis, we mainly use the metallicities of GCs measured from the \hbeta versus \mgfep grid
 because the main concerns in this study are the ages rather than the alpha-elements of GCs.

Panel (d) illustrates a comparison
  of the metallicities obtained from the grid method and the $\chi^2$ minimization method,
  which shows a good agreement
  between the two measurements. 
  We derive a linear least-squares fit:
  $\textrm{[Z/H]}{\rm ( \chi^2 )}
  =0.946(\pm0.073)\textrm{[Z/H]}{\rm ( grid )}-0.092(\pm0.052)$ with $rms=0.209$.
%
Panel (e) shows a comparison
  of the ages obtained from the grid method and the $\chi^2$ minimization method.
Panel (e) shows a reasonable agreement   between the two measurements with a large scatter.   We derive a linear least-squares fit:
  $\textrm{Age}{\rm (\chi^2 )}
  =1.017(\pm0.085)\textrm{Age}{\rm ( grid )}+0.057(\pm0.739)$ with $rms=2.942$.
Note that there are six GCs with ages younger than 5 Gyr derived from both methods. 
The five out of six GCs have relatively higher S/N ($>20$). 
Therefore, the existence of young GCs is not simply due to the low S/N spectra (see next section for detailed discussion).
%
Panel (f) shows a comparison
  of the \afe obtained from the grid method and the $\chi^2$ minimization method.
 It shows a reasonable agreement   between the two measurements with a large scatter.   We derive a linear least-squares fit:
  $\textrm{\afe}{\rm ( 
  \chi^2 )}=0.934(\pm0.064)\textrm{\afe}{\rm ( grid )}-0.042(\pm0.028)$ with $rms=0.125$.

\subsection{Distributions of Metallicity, Age and [$\alpha$/Fe]}\label{result-ama}

\begin{figure}
\plotone{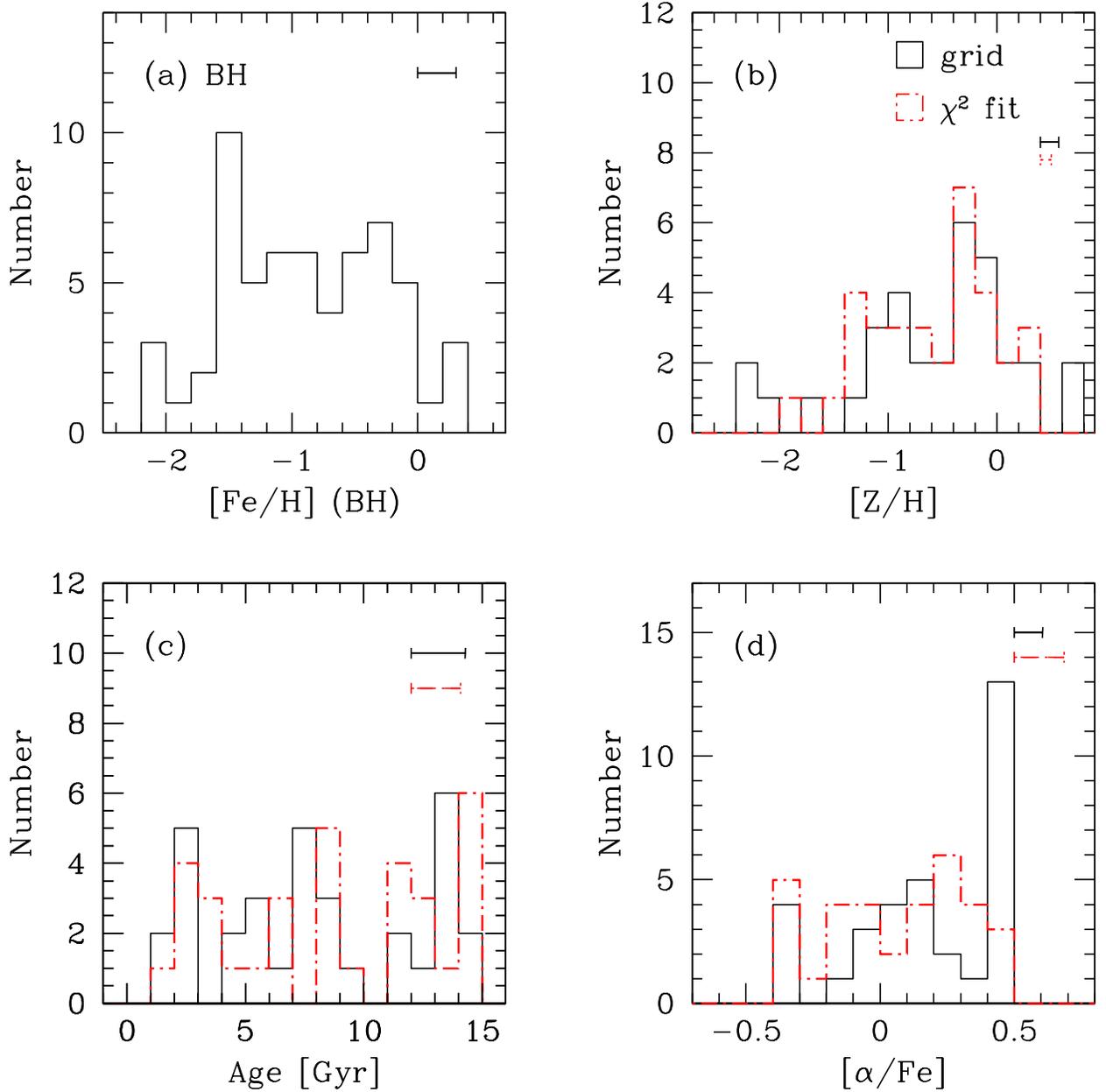}
\caption{ 
Metallicity, age, and [$\alpha$/Fe] distribution for NGC 4636 GCs.
(a) Distribution of [Fe/H] from the BH method.
(b) [Z/H] distributions obtained from \hbeta versus \mgfep grid.
(c) Age distribution. 
(d) \afe distribution.
 The solid and dot-dashed histograms represent the  distributions derived with grid and $\chi^2$ method, respectively.
 The error bars in the upper right of each panel indicate the mean error of each value.
\label{fig-numdist}}
\end{figure}

Figure \ref{fig-numdist} represents the distributions of 
  metallicity, age, and [$\alpha$/Fe] for the GCs in NGC 4636
  derived from the three methods.
Panel (a) shows that
  the [Fe/H] obtained from the BH method has a wide distribution at $-2.2 <$ [Fe/H] $< 0.4$
  with the mean value of [Fe/H] $= -0.90 \pm 0.12$. 
Two peaks appear to be at [Fe/H] $\sim -1.5$ and $\sim -0.4$.
The [Z/H] distribution obtained from the \mgfep grid method (Figure \ref{fig-numdist} (b)) 
  shows a similar wide range of $-2.4<$ [Z/H] $<0.8$ with the mean value of $-0.58 \pm 0.75$. 
  Two peaks are seen at $-1.0$ and $-0.2$.
The results from the $\chi^2$ minimization method (dot-dashed line) are similar
to those from the grid method (solid line).

\begin{figure}
\epsscale{1.0}
\plotone{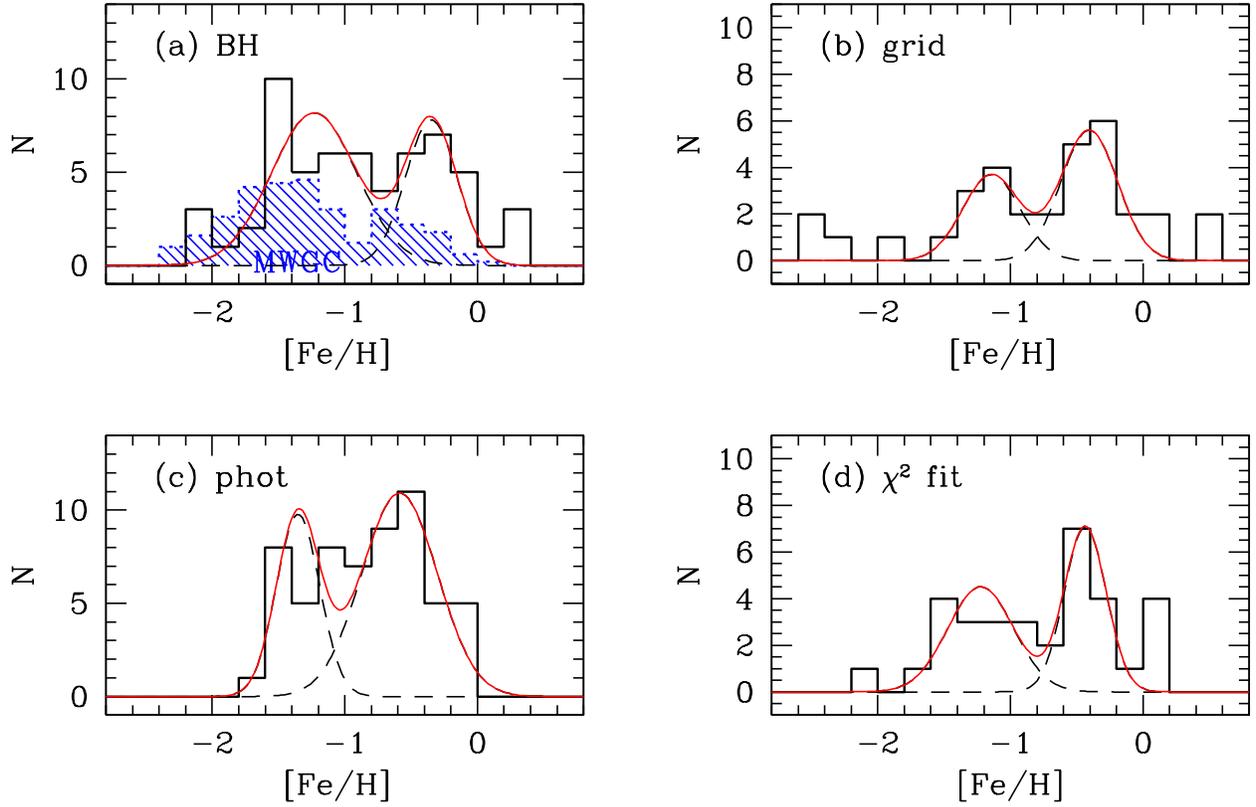}
\caption{ 
Metallicity distribution of NGC 4636 GCs.
(a) [Fe/H] distribution from the BH method. 
The hatched histogram represents the [Fe/H] distribution of the MW GCs \citep{har96}. 
The histograms in (b) and (d) represent [Fe/H] distributions obtained from the grid and $\chi^2$ minimization method, respectively.
(c) Photometric metellicity distribution obtained from $(C-T_1)_0$ by \citet{par12}
 and the double linear equation by \citet{lee08a}.
The curved solid lines represent double Gaussian fits.
\label{fig-metaldist}}
\end{figure}

We performed the KMM test to analyze quantitatively the bimodality of
   the metallicity distribution of NGC 4636 GCs.
The KMM test based on the algorithm of \citet{ash94} 
  gives an estimate of the improvement of the $n$-group fit over a single Gaussian.
In our sample,
the hypothesis of a unimodal metallicity distribution can be rejected 
    at the 97.1\%, 71.4\%, 98.4\%, and 95.2\% confidence level
  in the case of the [Fe/H] obtained  from the BH method, the \mgfep grid,  
    the color-metallicity relation, 
    and the $\chi^2$ minimization method, 
    respectively (see Figure \ref{fig-metaldist}).
These suggest that the metallicity distribution of NGC 4636 GCs  is marginally bimodal.

Figure \ref{fig-metaldist} represents the metallicity distributions 
  and their Gaussian fits determined from the KMM test. 
Panel (a) shows that two peaks are 
  at $\textrm{[Fe/H]}=-1.23 (\sigma=0.32)$, and $-0.35 (\sigma=0.19)$.
The distribution of [Fe/H] determined with \mgfep grid (panel (b))
  shows two peaks at
  $\textrm{[Fe/H]}=-1.14 (\sigma=0.22)$ and $-0.41 (\sigma=0.21)$. 
  The distribution of [Fe/H] from the $\chi^2$ minimization  method (panel (d))
  also shows two peaks at
  $\textrm{[Fe/H]}=-1.22 (\sigma=0.25)$ and $-0.44 (\sigma=0.16)$. 
In addition we derived the photometric metallicity distribution from the $(C-T_1)_0$ color \citep{par12} using  
  the double linear equation given in \citet{lee08a} (panel (c)). 
  It shows two peaks at 
 $\textrm{[Fe/H]}=-1.35 (\sigma=0.16)$ and $-0.59 (\sigma=0.28)$.
We also plot the metallicity distribution of the MW GCs 
  from \citet{har96} (2010 version) in panel (a) for comparison. 
It also shows  two peaks at 
 $\textrm{[Fe/H]}=-1.52 (\sigma=0.39)$ and $-0.52 (\sigma=0.23)$, each of which is at slightly lower values than the NGC 4636 GCs.

Figure \ref{fig-numdist} (c) shows that the age distribution of the NGC 4636 GCs 
  has a wide range from young (2 Gyr) to old (15 Gyr)
  with the mean grid age of 7.9 $\pm$ 0.8 Gyr and
  the mean $\chi^2$ fit age of 8.7 $\pm$ 2.0 Gyr.
 The results from both methods are similar. 
It seems that 
 there are three peaks at 3, 8, and 14 Gyr.
If we divide the age distribution into three bins of
$t<5$ Gyr, $5<t<10$ Gyr, and $t>10$ Gyr,  
  the number ratios are 9:13:11 (27:39:34\%) in the grid age
  and 9:10:14 (27:30:43\%) in the $\chi^2$ fit age. 
The fraction of the young GCs with age $<5$ Gyr in our sample is about 27\%,
which is a slightly larger than that (18\%) in NGC 5128 \citep{woo10}.

To check the robustness of the large fraction of young GCs in NGC 4636,
  we first divide the GCs into two groups based on their S/N: 
  22 GCs with S/N $>20$ and 11 GCs with S/N $<20$.
  The GCs in the two groups do not show any difference in their age distributions. 
  There are six young (age $< 5$ Gyrs) GCs among 22 GCs with S/N $>20$. 
  Similarly, there are three young GCs among 11 GCs with S/N $<20$. 
  These suggest that the existence of young GCs is not simply because of spectra with low S/N. 
However, Brodie et al. (2005) argued that young or intermediate GCs may be 
due to relatively low S/N spectra or to sky subtraction difficulties. 
We also computed the probability of the existence of young GC with age $<5$ Gyr and S/N $>20$ 
  using a simple simulation with GCs of age = 9 Gyr (a mean age of NGC 4636 GCs) 
  and of Hbeta error = 0.34 A (a mean value of measured errors). 
The fraction of young GCs with age $<5$ Gyr derived from simulations is 12\%, 
which is 
 smaller than the observational fraction (27\%).
This result indicates that the young GCs in NGC 4636 could be caused in part by observational uncertainty. 
The reliability of the existence of young GCs needs to be checked with better data in the future.

Alpha-element abundances indicate 
  the duration of star formation in a GC.
The GCs with an enhanced \afe ratio 
  experienced rapid star formation over less than 1 Gyr timescale,
  while solar \afe GCs have an extended star formation history \citep{bro06}. 
Figure \ref{fig-numdist} (d) displays the distribution of \afee.
  The overall distribution in the grid method shows a single peak with a large dispersion 
  except for those at both ends,  
  while that from the $\chi^2$ minimization method does not show a dominant peak.
The mean \afe is derived to be  $0.20 \pm 0.11$ in the grid method 
  and $ 0.07 \pm 0.05$ in the $\chi^2$ minimization method. 
This mean value  is between the values for the NGC 5128 GCs 
 (\afe $= 0.13 \pm 0.03$ in the grid method  
  and \afe $= 0.06 \pm 0.03$ in the $\chi^2$ minimization method,
  see Table \ref{tab-alphadist})
  and for the MW GCs (\afe $= 0.36 \pm 0.01$).
%
%
This suggests that the formation period of NGC 4636 GCs might have been more extended, 
  as in the case of NGC 5128 GCs, than the MW GCs.

\subsection{Radial distribution}

\begin{figure}
\epsscale{0.8}
\plotone{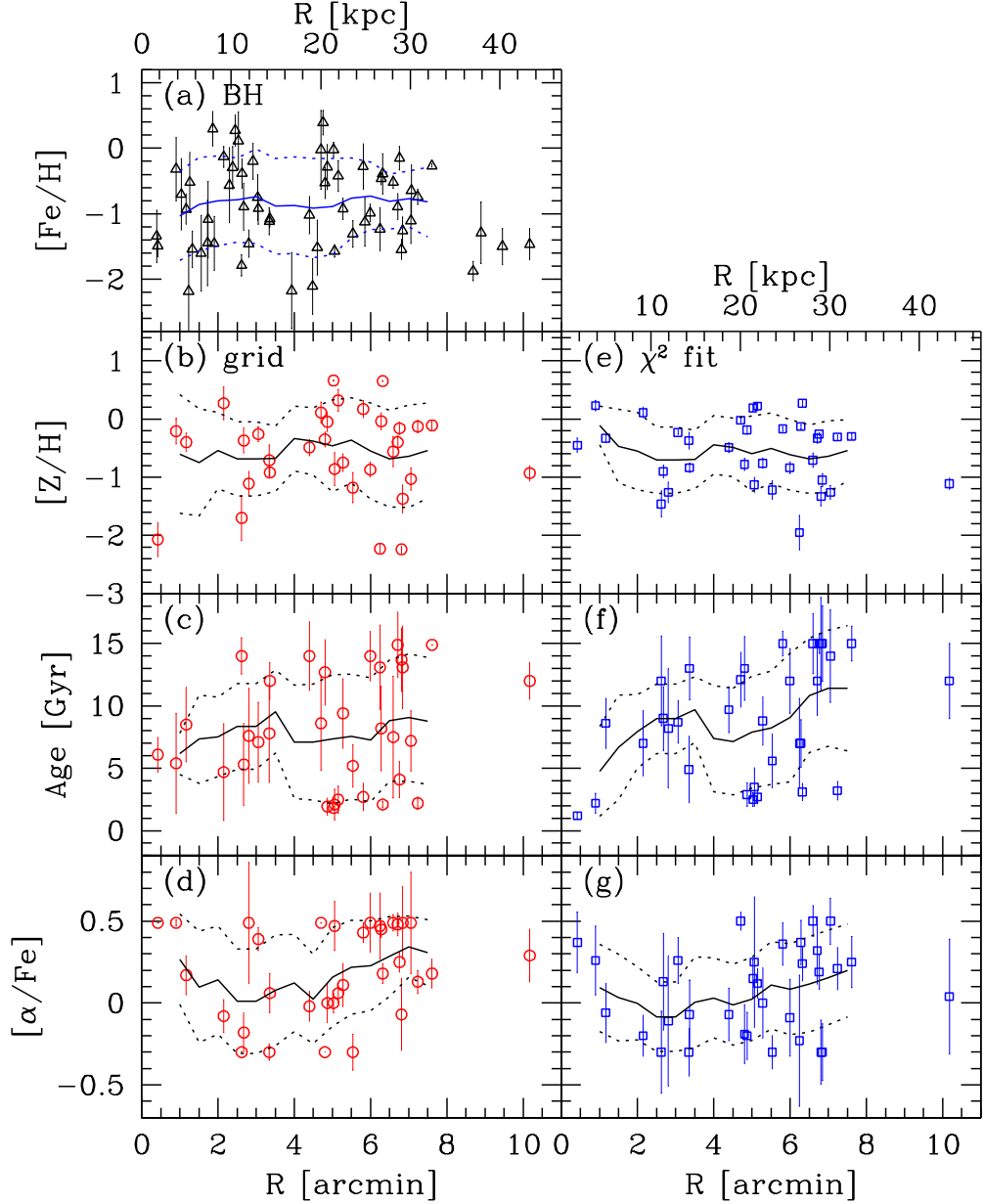}
\caption{ 
Radial variation of metallicity, age, and \afee:
(a) [Fe/H] from the BH method,
(b) [Z/H], (c) Age, and (d) \afe from the grid method, and
(e) [Z/H], (f) Age, and (g) \afe from the $\chi^2$ method.
The solid and dotted lines represent the mean values and their dispersions 
derived using a moving radial bin (with 2.5 arcmin width and 0.5 arcmin step).
\label{fig-radidist}}
\end{figure}

We investigated  the distributions of
  the metallicity, age, and [$\alpha$/Fe] of the NGC 4636 GCs  
  as a function of the galactocentric distance.
Figure \ref{fig-radidist} (a), (b), and (e) 
  show that  [Fe/H] and [Z/H] (from BH, grid, and $\chi^2$ minimization methods) vary little with 
  the galactocentric radii at $R<8\arcmin$ ($<34$ kpc), but show a large scatter. 
 It is noted that four GCs at  $8\arcmin <R<10\arcmin.5$ have on average $\sim 0.5$ dex lower [Fe/H] values than the GCs in the inner region. 
%
%
Figure \ref{fig-radidist} (c) and (f) show that 
  the mean age of the NGC 4636 GCs changes little with 
  the galactocentric radii 
  and the dispersion in age is large. 
This trend is similarly found in M60 GCs \citep{pie06}. 
%
Panels (d) and (g) 
  show no radial dependence of \afee. 

\subsection{Relation between metallicity and age}

\begin{figure}
\epsscale{0.70}
\plotone{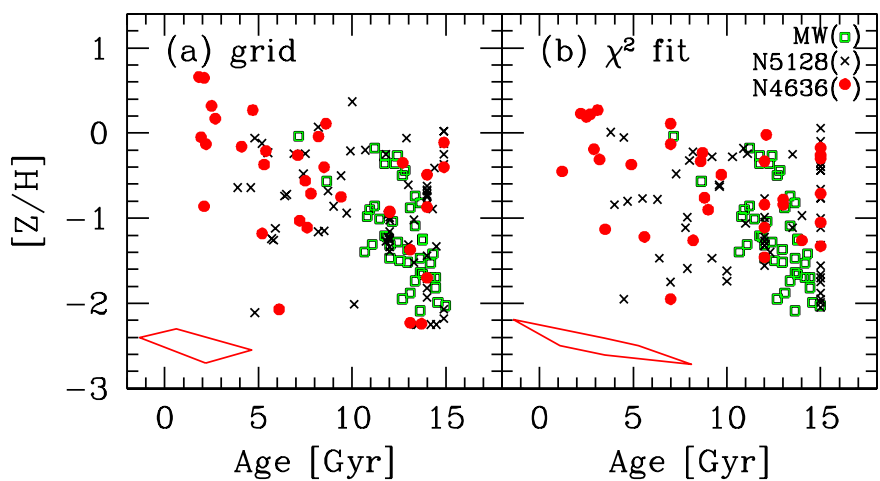}
\caption{ 
Relations between metallicity and age 
from the grid (a) and the $\chi^2$ minimization method (b).
The circles and crosses represent the NGC 4636 GCs in this study 
  and the NGC 5128 GCs derived from indices in \citet{woo10}, respectively.
The squares represent the MW GCs by \citet{har96} and \citet{car10}. 
The large boxes represent the mean errors for NGC 4636 GCs.
\label{fig-relama}}
\end{figure}

Figure \ref{fig-relama} 
displays the relations between metallicity and age for 
the NGC 4636 GCs in comparison with 
  NGC 5128 GCs and the MW GCs. 
The NGC 4636 GCs show, on average, an anti-correlation between age and metallicity.
Younger GCs in NGC 4636 have higher metal abundances in both grid and $\chi^2$ minimization results.
 It is also seen in the case of NGC 5128 GCs (see also \citet{woo10}) and the MW GCs.
This correlation is also seen for the GCs in M60 \citep{pie06}
 and M86 \citep{par12b}. 
Note that this correlation could be caused by the correlated errors of age and metallicity 
  (see the error boxes in Figure \ref{fig-relama}) \citep{tra00,kun01,pau10} 
  or by a biased sample selection (e.g., extremely bright GCs) \citep{woo10}.
 The age of the youngest GCs in the MW is 7 Gyr, 
 but NGC 4636 and NGC 5128 have a significant number of  GCs younger than 7 Gyr.
The relative fraction of old ($>10$ Gyr) GCs in NGC 4636 is smaller than that of NGC 5128 GCs and much smaller than that of MW GCs. 

\section{Discussion}\label{discuss}

In this study, 
we derived the metallicity, age, and \afe for the GCs in NGC 4636 from the spectroscopic data.
Here we compare these with those of GCs in other nearby gEs
and discuss the results in the context of 
GC formation in gEs.
We found six gEs for which the spectroscopic data of their GCs are available in the literature.
  We compiled the Lick line index measurements for the GCs in each galaxy from the following references:
\citet{coh98} for M87,
\citet{bea00} and \citet{coh03} for M49, 
\citet{woo10} for NGC 5128,
\citet{pie06} for M60,
\citet{cen07} for NGC 1407, and
\citet{for01} for NGC 1399.
To have a fair comparison of physical parameters among gEs, 
  we re-derived the metallicity, age, and \afe for the GCs in each gE from the line indices 
  using both grid and $\chi^2$ minimization methods as done for NGC 4636 GCs (see Section 3.2).  
We also compared the GCs in gEs with the MW GCs.
The data of the MW GCs 
  are obtained
  from \citet{har96} (2010 version) for metallicity,
  from \citet{mar09} and \citet{dot10} for age, and
  from \citet{car10} for \afee.
  
\subsection{Metallicity distribution of GCs in gEs}

The color bimodality in recent extragalactic GC studies is
  one of the important ingredients 
  for GC formation scenarios \citep{bro06}. 
This bimodal color distribution is interpreted as 
  an intrinsic property due to a bimodal metallicity distribution \citep{bro06, spi08, woo10, chi12},
  or an apparent feature simply due to a nonlinear color-metallicity relation \citep{yoo06,yoo11a,yoo11b}.
It is not yet clear which of these two is more supported by observational results.
Here we investigate  the metallicity distributions of the GCs in seven gEs.

\begin{figure}
\epsscale{1.0}
\plotone{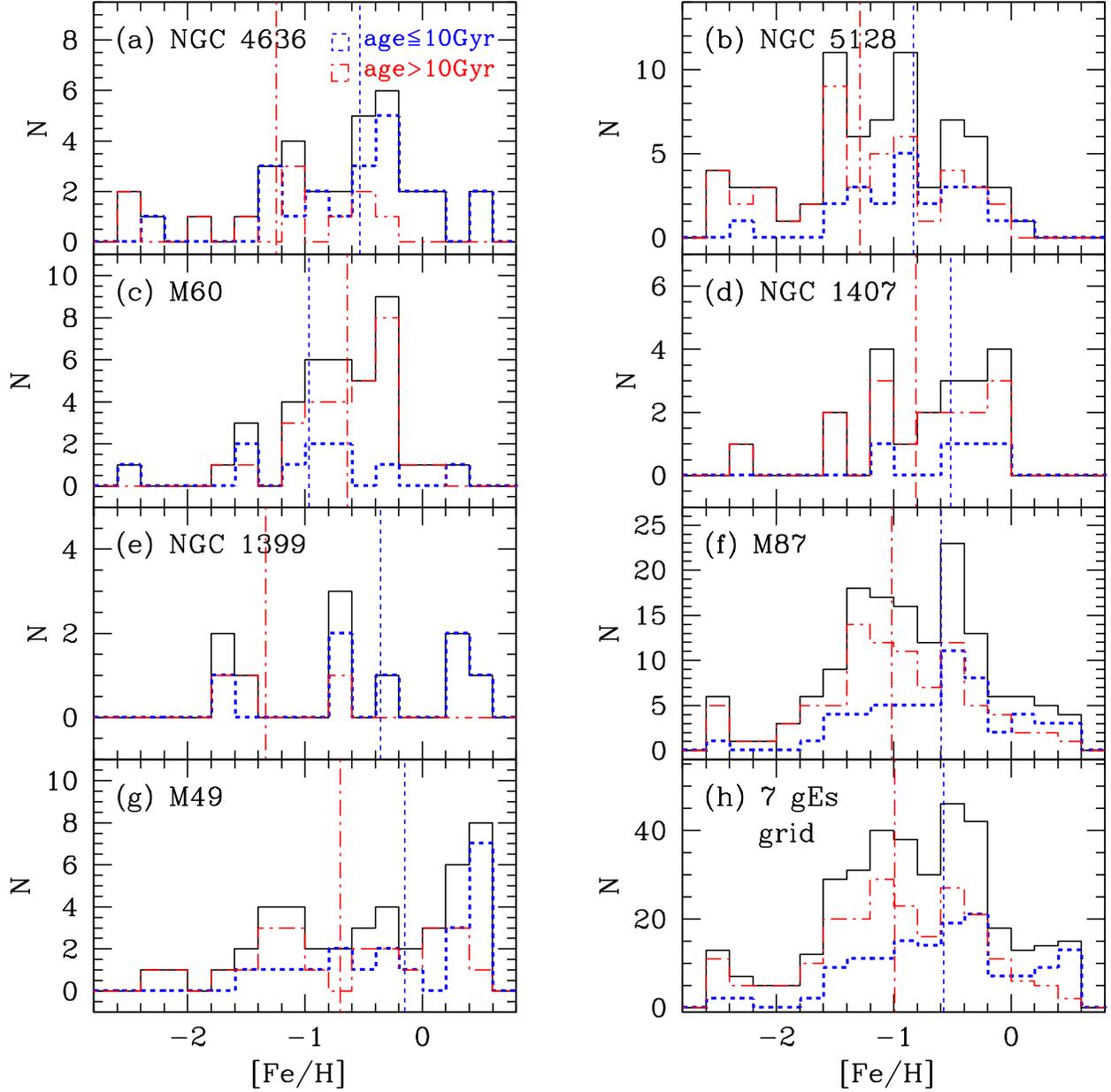}
\caption{ 
Metallicity distribution of GCs in gEs from the grid method.
The solid, dot-dashed, and dotted line histograms for gEs show the [Fe/H] distributions of all GCs, old GCs ($>$ 10 Gyr), and young GCs ($\le$ 10 Gyr), respectively.
The vertical dotted and dot-dashed lines represent the mean metallicities for young and old GCs, respectively.
In panel (h)
the histograms represent the metallicity distribution
  for the combined sample of GCs in all seven gEs.  
\label{fig-gridgEsmetal}} 
\end{figure}

Figure \ref{fig-gridgEsmetal} displays the  distributions of metallicity derived with the grid method for the GCs in seven gEs. 
The metallicity distributions of GCs in most gEs 
 seem to be bimodal 
 as in the case of the MW GC system (see Figure \ref{fig-metaldist} (a)).
We run the KMM test for quantitative analysis of the metallicity
distribution, and list the results for the Gaussian peaks and their confidence levels in Table \ref{tab-metalpeak}.
There is no strong 
  evidence for the bimodality for 
  NGC 4636, M60, NGC 1407, and NGC 1399 that have a relatively small number of GCs. 
  However, the bimodality for  NGC 5128, M87, and M49 is supported by 
  the KMM test with a confidence level $\gtrsim$ 98\%. 

We combine the data derived from the grid method for all the GCs in seven gEs,
and show the resulting metallicity distribution in Figure \ref{fig-gridgEsmetal} (h).
The combined gE sample  (304 GCs) shows clearly 
  a bimodal distribution (155 MP GCs and 149 MR GCs). The KMM test yields two Gaussian peaks at 
  $\textrm{[Fe/H]}=-1.21 (\sigma=0.32)$, and $-0.42 (\sigma=0.27)$    with 99.9\% confidence level. 
 The MR peak for the combined gE sample is
 $\sim 0.1$ dex higher than that for the MW GCs ( $\textrm{[Fe/H]}=-0.52  (\sigma=0.23)$ ), 
 while the MP peak is $\sim 0.3$ dex higher than that of the MW GCs 
 ($\textrm{[Fe/H]}=-1.52  (\sigma=0.39)$).  
 It is noted that   there are two more minor components at both ends 
 ([Fe/H] $< -2.0$ and [Fe/H] $> 0.0$) in the combined gE sample.

\begin{figure}
\epsscale{1.0}
\plotone{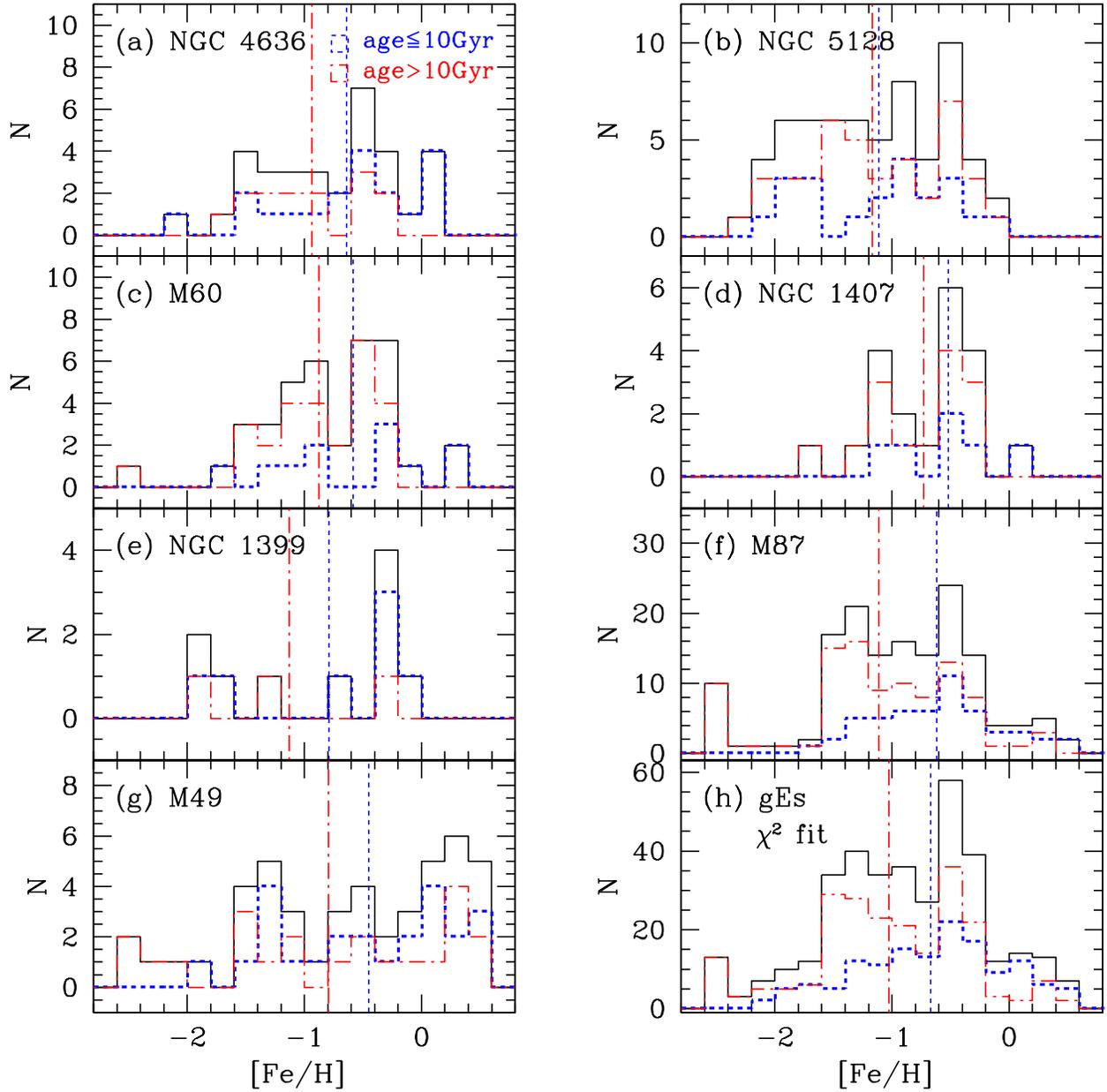}
\caption{ 
Metallicity distribution of GCs in gEs from the $\chi^2$ minimization method.
Symbols are same as Figure \ref{fig-gridgEsmetal}.
\label{fig-chi2gEsmetal}} 
\end{figure}

Figure \ref{fig-chi2gEsmetal} displays the  distributions of metallicity derived with the $\chi^2$ method for the GCs in seven gEs. 
The KMM test results for the Gaussian peaks and their confidence levels are given in Table \ref{tab-metalpeak}.
There is no strong evidence for the bimodality only for two gEs (NGC 1399 and NGC 1407).
  However, the bimodality for NGC 4636,  NGC 5128, M60, M87, and M49 is supported by 
  the KMM test with a confidence level $\ge$ 95\%.
The combined gE sample  (315 GCs) shows clearly 
  a bimodal distribution (167 MP GCs and 148 MR GCs). The KMM test yields two Gaussian peaks at 
  $\textrm{[Fe/H]}=-1.25 (\sigma=0.32)$, and $-0.42 (\sigma=0.25)$    with 99.9\% confidence level,
  which are very similar to those from the grid method, 
  $\textrm{[Fe/H]}=-1.21 (\sigma=0.32)$, and $-0.42 (\sigma=0.27)$. 
 It is noted that   there are two more minor components at both ends 
 ([Fe/H] $< -2.0$ and [Fe/H] $> 0.0$) in the combined gE sample, as found in the results
 from the grid method.

To examine any age dependence of the metallicity distribution of the GCs in gEs, 
we divide the GC samples into two groups: 
young GCs with age $\le$ 10 Gyr   and old GCs with age $>$ 10 Gyr.
In Figures \ref{fig-gridgEsmetal} and \ref{fig-chi2gEsmetal},  
  we plot the
   metallicity distribution of the young GCs and  old GCs (the dotted  and dot-dashed lines, respectively). 
We also derive the mean metallicities of the young GCs and old GCs,
 which are marked with the vertical lines in Figures \ref{fig-gridgEsmetal} and \ref{fig-chi2gEsmetal} (see also Table \ref{tab-metaldist}).
The mean values are determined with the biweight location \citep{bee90}, and
their errors indicate the 68\% confidence levels obtained with a bootstrip method.
%
The mean metallicity of all GCs  in gEs ([Fe/H] $= -0.80\pm0.04$ and  $ -0.86\pm0.04$ 
from the grid method and the $\chi^2$ minimization method, respectively) is much higher than
that for the MW GCs ([Fe/H] $= -1.28\pm0.05$). 
The mean metallicity of the young GCs in each gE and the MW is 0.2 -- 1.3 dex higher
than that of the old GCs. 
For the combined sample, 
the mean metallicities of the young GCs and old GCs are 
[Fe/H] $= -0.55\pm0.05$ and [Fe/H] $= -0.96\pm0.05$ in the case of the grid  method,
and [Fe/H] $= -0.65\pm0.06$ and [Fe/H] $= -1.00\pm0.04$ in the case of the $\chi^2$ minimization  method.
Thus the metallicity of the young GCs is, on average, $\sim0.4$ dex higher than that of the old GCs.
It is noted that both old and young GCs appear to show a bimodal distribution.
  However, the MP component is dominant for the old GCs,
  while the MR component is dominant for the young GCs.

We also examine a relation between metallicity and \afee.
We divide the sample into two groups: high \afe $> 0.2$ and low \afe $\le 0.2$.
 The results are listed in Table \ref{tab-metaldist}. 
The mean metallicity of GCs does not show any significant dependence on \afee, 
  except for two gEs (NGC 1407 and NGC 1399) with a small number of GCs. 
%
Note that some of the differences between galaxies could be dominated 
  by selection effects (e.g., different brightness of GCs) or by small number statistics.

\subsection{Age distribution of GCs in gEs}

\begin{figure}
\epsscale{1.0}
\plotone{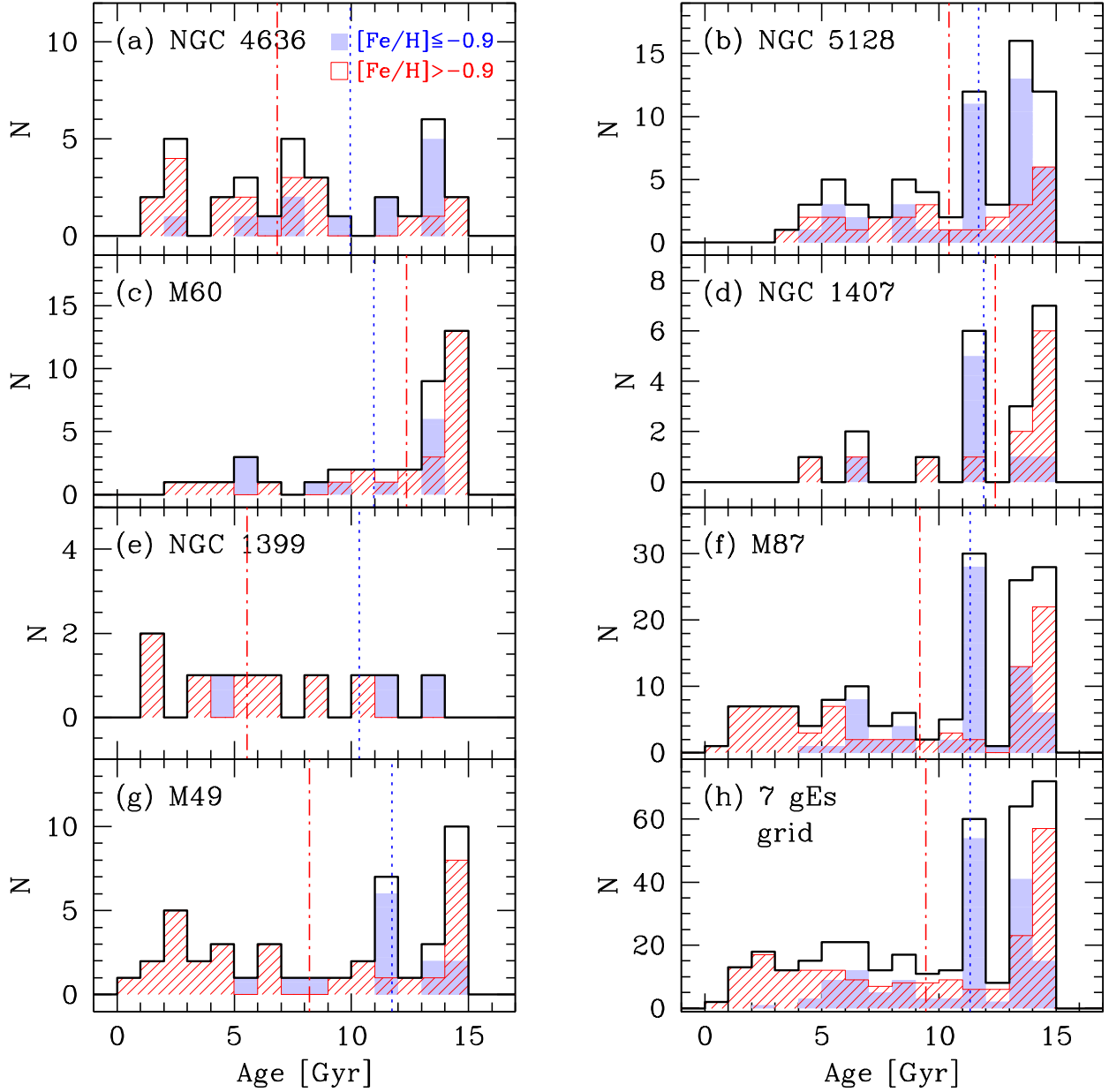}
\caption{ 
Age distribution of the MP ([Fe/H] $\le -0.9$) GCs (shaded histograms) 
 and the MR ([Fe/H] $> -0.9$) GCs (hashed histograms) in gEs 
 derived from the grid method.
The solid line histograms represent the sum of the MP and MR groups.
The mean ages of the MP and MR groups are shown by 
  the dotted and dot-dashed lines, respectively.
The histograms on panel (h) show the age distributions for the combined sample of GCs in seven gEs. 
\label{fig-gridgEsagedfeh}}
\end{figure}

\begin{figure}
\epsscale{1.0}
\plotone{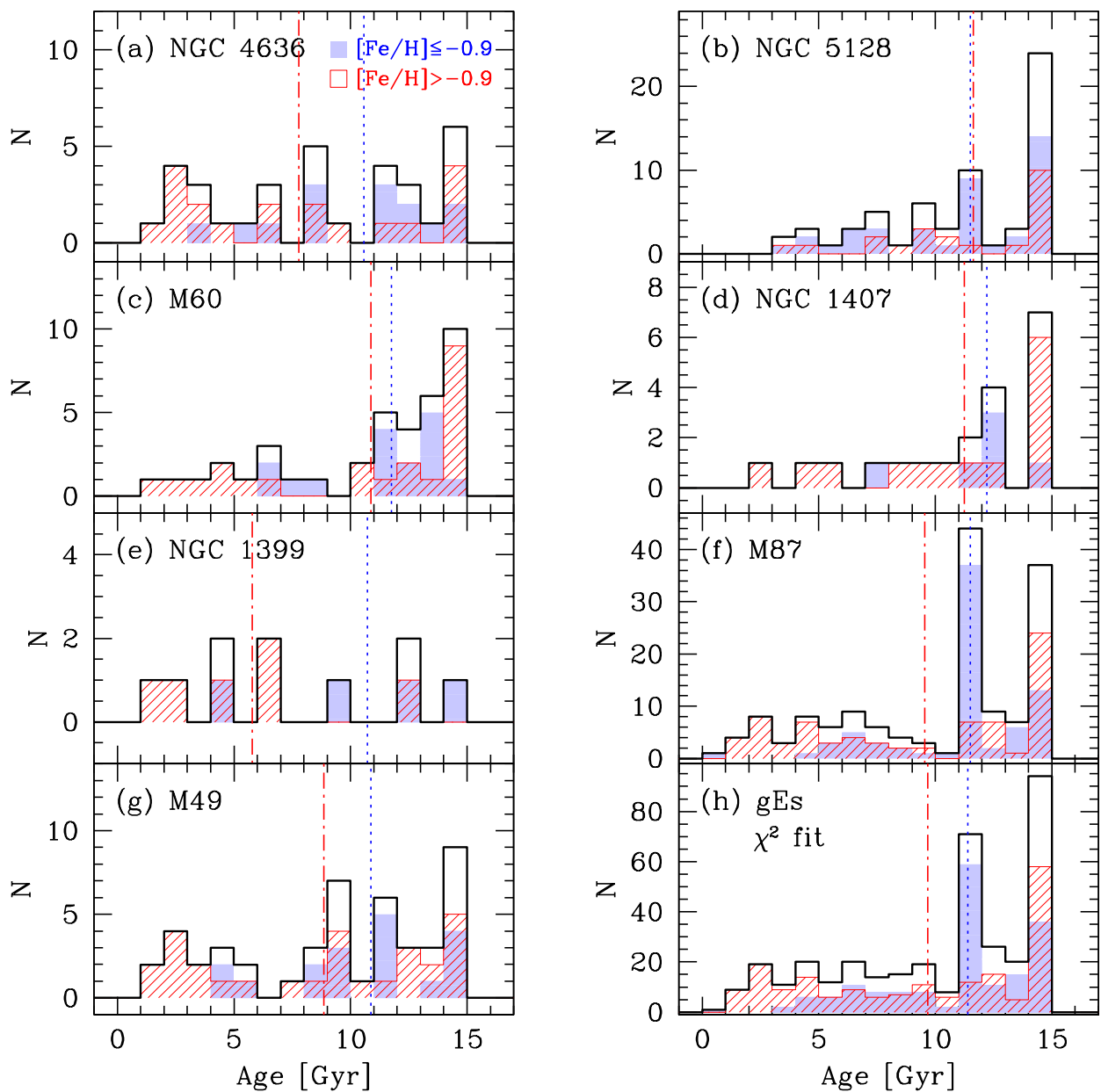}
\caption{ 
Age distribution of GCs
 in gEs derived from the $\chi^2$ minimization method.
Symbols are same as Figure \ref{fig-gridgEsagedfeh}.
\label{fig-chi2gEsagedfeh}}
\end{figure}

Figures \ref{fig-gridgEsagedfeh} and  \ref{fig-chi2gEsagedfeh} show the age distributions of the GCs 
in gEs derived from the grid method and from the $\chi^2$ minimization method, respectively.
We also plot the age distribution of the GCs in the combined sample for gEs (panel (h)). 
We divide the sample into subgroups according to their metallicity
and \afee: MP groups with [Fe/H] $\le -0.9$ and MR groups with [Fe/H] $>-0.9$, 
and low \afe groups with \afe $\le 0.2$ and high \afe groups with \afe $>0.2$.
We derive the mean ages of each group, which are listed in Table \ref{tab-agedist}.
The mean values of the metallicity subgroups are shown by the vertical lines in Figures \ref{fig-gridgEsagedfeh} and \ref{fig-chi2gEsagedfeh}. 

\begin{figure}
\epsscale{1.0}
\plotone{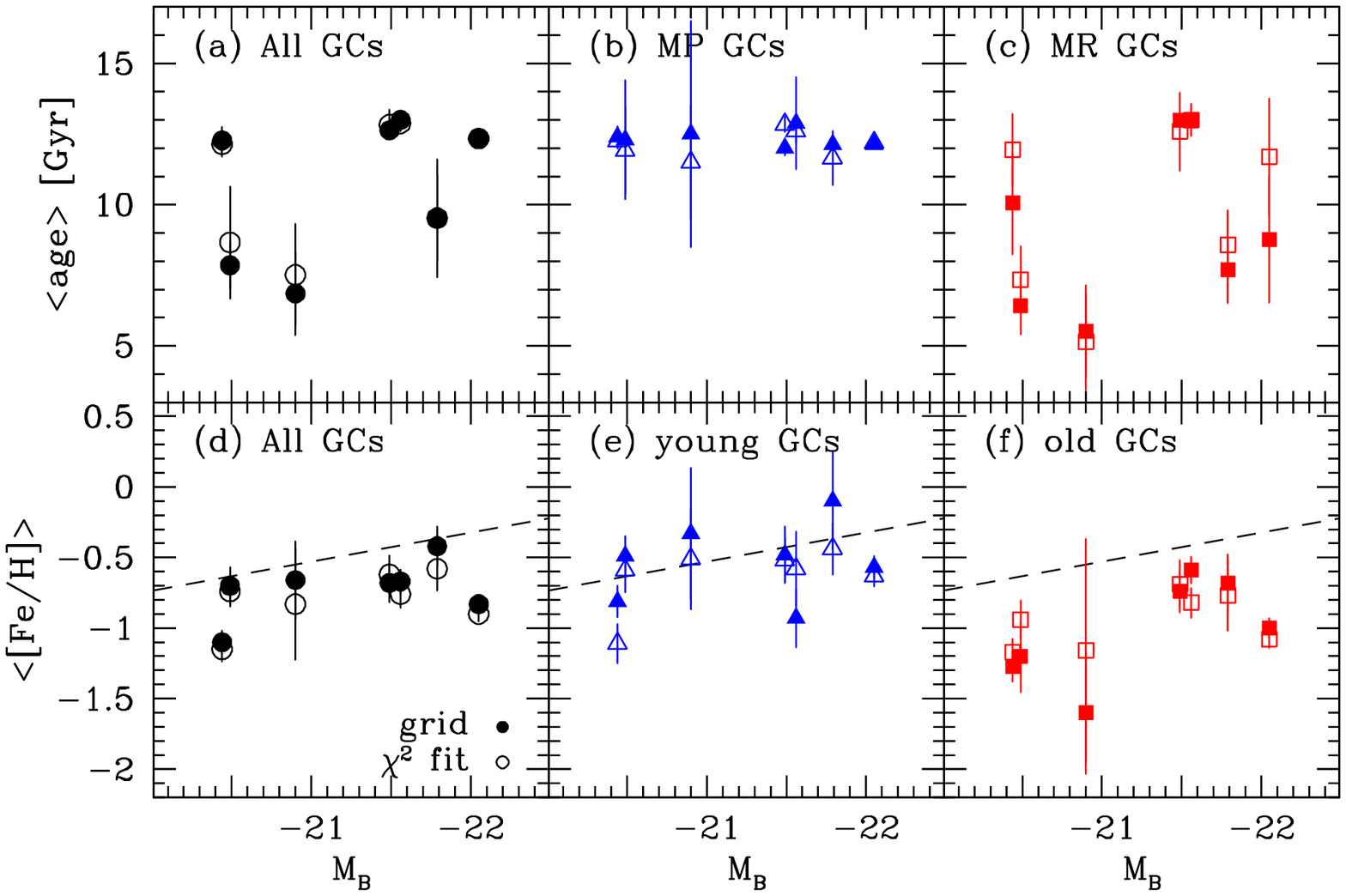}
\caption{ 
Mean ages and metallicities of GCs as a function of absolute magnitude ($M_{B}$) of gEs.
(a,d) All GCs,
(b) MP GCs ([Fe/H] $\le -0.9$), 
(c) MR GCs ([Fe/H] $> -0.9$),
(e) young GCs (age $\le 10$ Gyr), and
(f) old GCs (age $> 10$ Gyr).
$M_{B}$'s of gEs are obtained from HyperLeda \citep{pat03}. 
The filled and open symbols are from  the grid method and the $\chi^2$ minimization method, respectively.
The dashed lines represent a relation derived from photometry of GCs in early-type galaxies in Virgo by \citet{pen06}.
%
\label{fig-gEsmagage}} 
\end{figure}

Several features are noted in these figures.
First, all gEs in this sample show  a wide range of ages for GCs, from 1 Gyr to 15 Gyr. 
This is in stark contrast to the case of the MW GCs. The MW GCs are all older than 7 Gyr and mostly older than 10 Gyrs (see Figure \ref{fig-relama}).
Second, the mean age for the MP GCs in the combined sample ($12.2\pm0.1$ Gyr from both methods) 
is about 3 Gyr larger than that for the MR GCs ($9.2\pm0.7$ Gyr from the grid method
and $9.7\pm0.7$ Gyr from the $\chi^2$ minimization method, respectively).
The MW GCs show similar trends, but with  a smaller difference of
1 Gyr:
$12.9\pm0.3$ Gyr for the MP GCs, $11.8\pm0.2$ Gyr for the MR GCs.
Third, the mean age for the high \afe GCs in the combined sample is similar to that for the low \afe GCs,  
while the high \afe GCs in the MW are, on average, older than  the low \afe GCs.
In Figure \ref{fig-gEsmagage}
  we plot the mean age and metallicity for all, MP, and MR GCs
  as a function of absolute magnitude ($M_{B}$) of gEs.
  The mean ages of the GCs (in the upper panels) do not show any systematic dependence on $M_{B}$ of gEs.
  The mean ages of the MP GCs show little scatter, while those of the MR GCs show a large scatter.
  It is noted, however, that two galaxies with low luminosity (NGC 1399 and NGC 4636)
  have the smallest mean ages of MR GCs.
The mean metallicities  of the GCs derived from both methods
(in the lower panels) also show little dependence
on the absolute magnitudes of their host galaxies.
Note that the results for all GCs and old GCs in gEs are not  consistent  with the results derived from photometry of GCs for early-type galaxies in Virgo by \citet{pen06} (dashed lines), while those for young GCs are. 
Further studies are needed to explain this difference.

\subsection{\afe distribution of GCs in gEs}

\begin{figure}
\epsscale{1.0}
\plotone{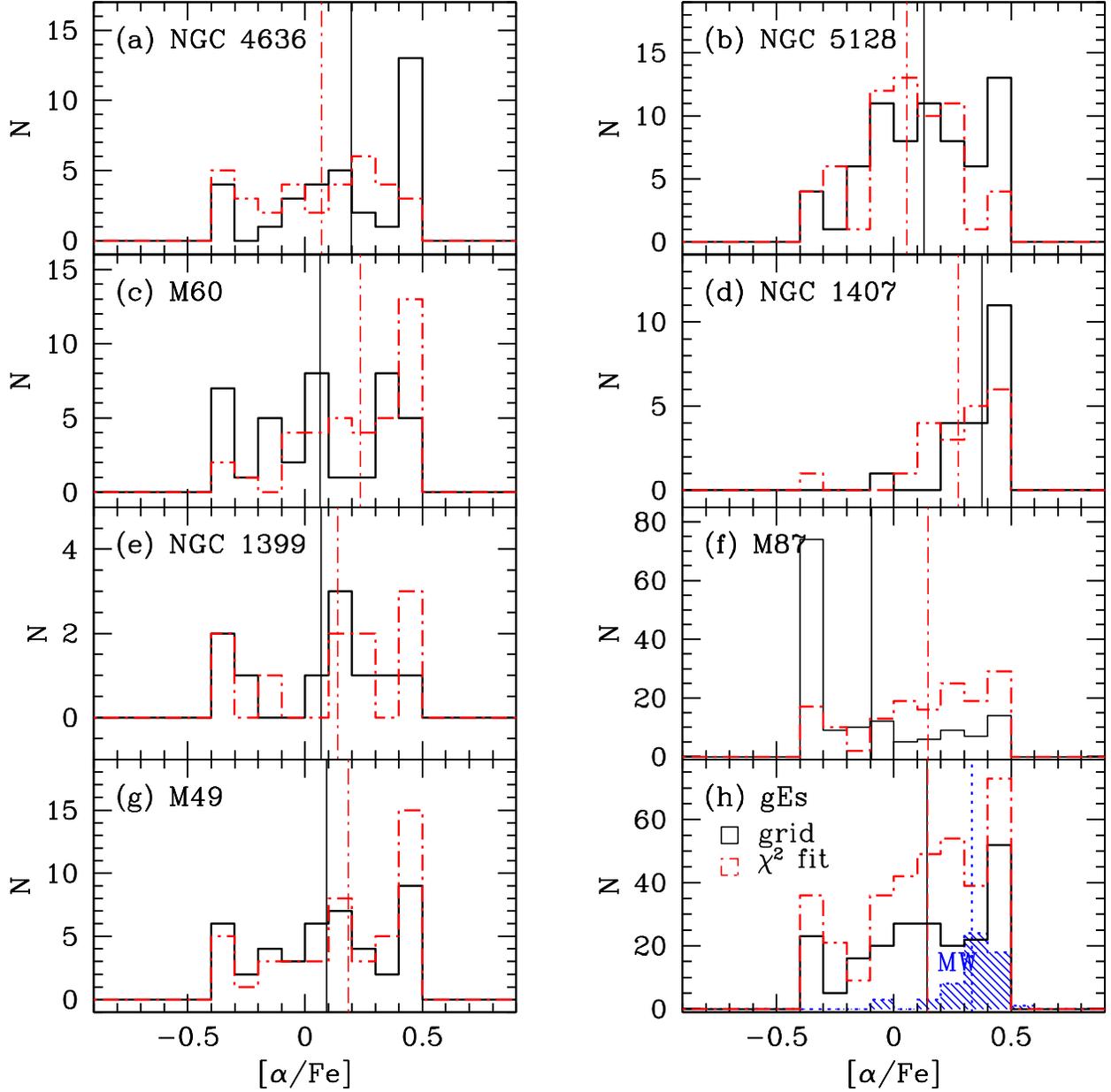}
\caption{ 
\afe distribution of GCs in gEs from the grid method (solid) and
the $\chi^2$ minimization method (dot-dashed).
The vertical lines represent 
 the mean values of the \afe of GCs in each galaxy.
Panel (h) shows the \afe distribution for the combined sample of GCs 
in six gEs (grid method) excluded M87 GCs and in seven gEs ($\chi^2$ fit).
The dotted histogram and line represent the MW GCs.
\label{fig-gEsafe}}
\end{figure}

Figure \ref{fig-gEsafe} shows the \afe distributions of the GCs in gEs. 
We also plot the \afe distribution of the GCs in the combined sample for gEs and the MW GCs in panel (h).
Here we exclude the \afe values of M87 GCs in the combined sample from the grid method
  because of abnormal value of \mgtwo Lick indices in the literature.
We divide the sample into subgroups according to their metallicity and age as done in the previous sections. 
We derive the mean \afe of each group, which are listed  in Table \ref{tab-alphadist}.
%
Notable features are as follows. 
First, 
  the mean value of \afe for the gE GCs (  \afe $= 0.14 \pm 0.02$ and $0.16\pm0.01$ from the grid method
  and the $\chi^2$ minimization method, respectively) is about 0.2 dex smaller than that for the MW GCs, 
  \afe $=0.36\pm0.01$. 
Second, 
  the mean values of \afe for the MP and MR GCs in gEs are similar.
Third, 
the mean values of \afe for the young and old GCs in gEs are also similar.
The old GCs in the MW have much higher \afee,
$0.35 \pm 0.11$ than the young GCs \afee,
$-0.02 \pm 0.02$, although the number of the young GCs is only two.

These \afe results of the GCs in gEs appear to be inconsistent with 
  those in seven early-type galaxies given by \citet{puz05b}, 
  who reported that the mean \afe for these gE GCs is high ($0.47\pm0.06$). 
  \citet{puz05} also found that the MR GCs exhibit lower \afe enhancements than the MP GCs, 
  and that \afe of GCs is independent from the age of GCs.
The low \afe for gE GCs in this study seems to be more consistent with the stellar populations 
  in massive elliptical galaxies expected from the dissipative hierarchical merging model \citep{tho99}. 
  The discrepancy between this study and \citet{puz05b} could be caused by 
  the mass difference in their host galaxies 
  because the our sample galaxies are more massive than those of \citet{puz05b}.

\subsection{Formation of GCs in gEs seen through the metallicities and ages}

\begin{figure}
\epsscale{1.0}
\plotone{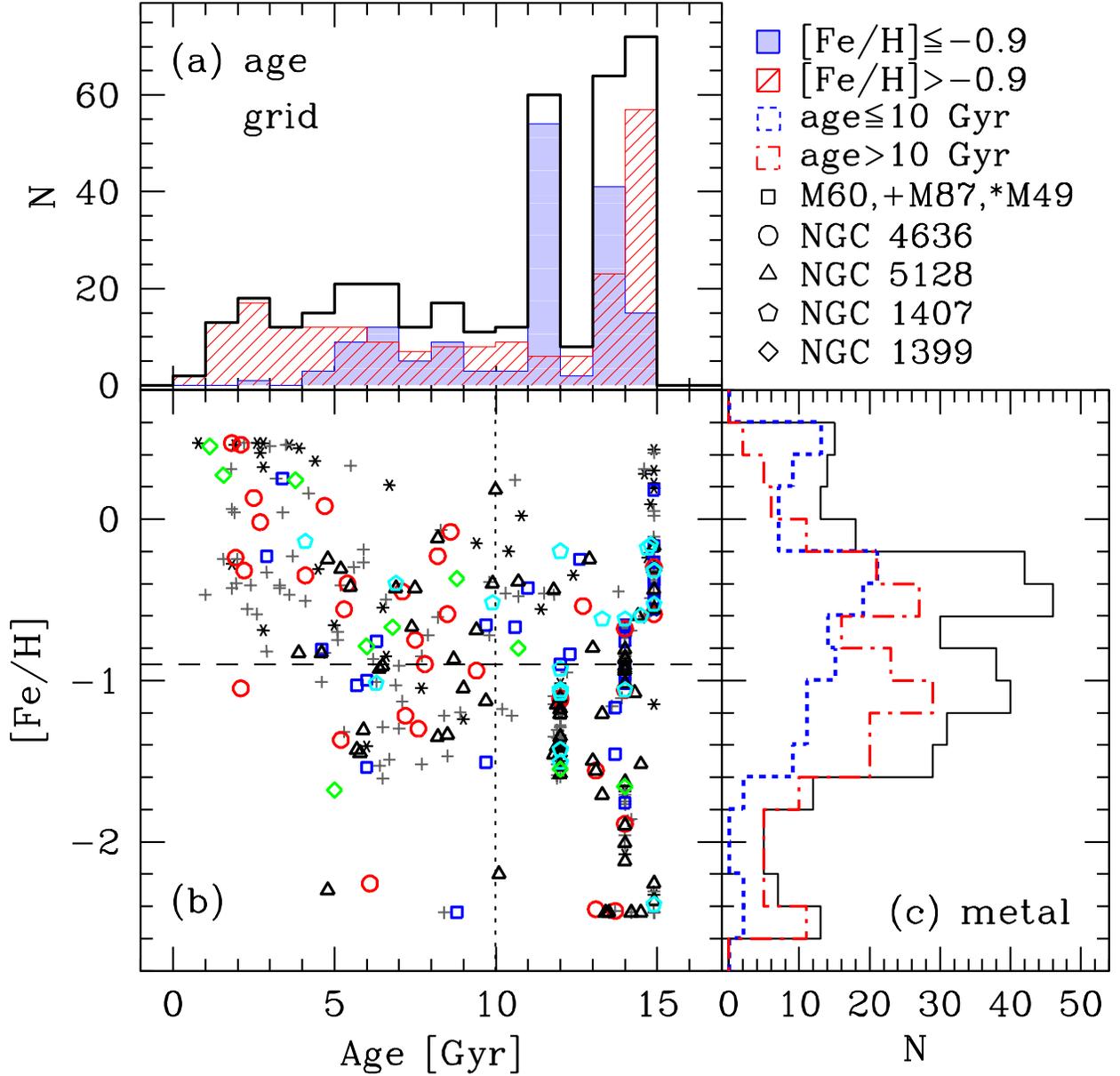}
\caption{ 
Age-metallicity relation of GCs in gEs from the grid method.
(a) The solid, shaded, and hashed histograms represent 
  the age distributions of all, MP, and MR GCs, respectively.
(c) The dotted and dot-dashed histograms show
  the [Fe/H] distributions for the combined sample of GCs with ages smaller and larger than 10 Gyr,  respectively, in seven gEs.
  The solid histogram represents the metallicity distribution of all GCs. 
(b) The squares, circles, triangles, pentagons, diamonds, plus, and stars
  represent the GCs of M60, NGC 4636, NGC 5128, NGC 1407, NGC 1399, M87, and M49, respectively.
The vertical  dotted line is an age criterion, 10 Gyr, 
  to divide young and old GCs
  and the horizontal dashed line is a metallicity criterion, [Fe/H] = $-0.9$ dex, 
  to divide MP and MR GCs. 
\label{fig-gridgEsagefeh}} 
\end{figure}

\begin{figure}
\epsscale{1.0}
\plotone{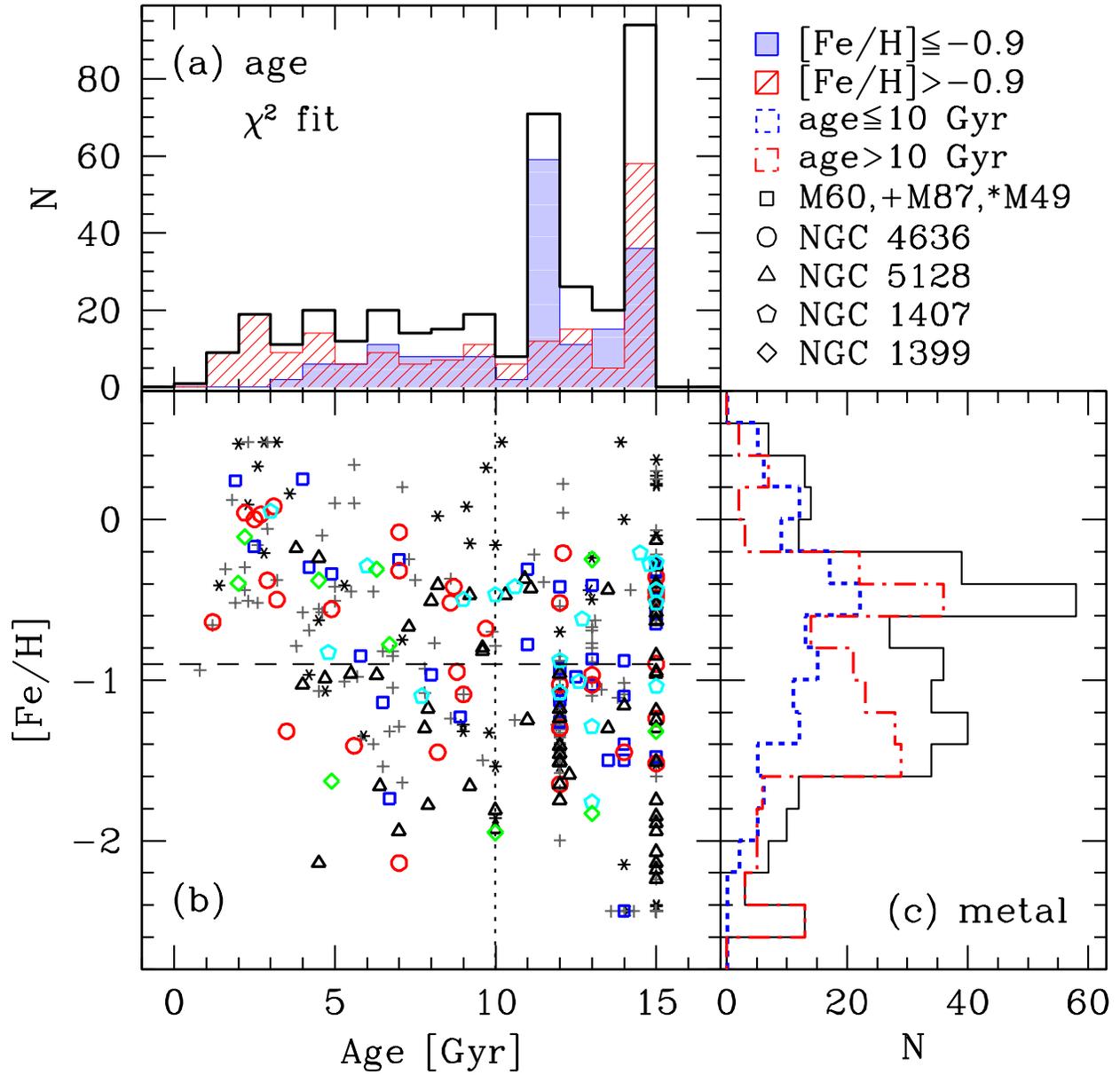}
\caption{ 
Age-metallicity relation of GCs in gEs from the $\chi^2$ minimization method.
Symbols are same as Figure \ref{fig-gridgEsagefeh}.
\label{fig-chi2gEsagefeh}} 
\end{figure}

In Figures \ref{fig-gridgEsagefeh} and \ref{fig-chi2gEsagefeh} we display the metallicity  and age of gE GCs
 derived from the grid and $\chi^2$ minimization methods:  
(a) age distributions for all, MP, and MR GCs,
(b) metallicity and age relation, and
(c) metallicity distributions for all, young, and old GCs.
The metallicities for gE GCs show an overall anti-correlation with ages, although there is a large scatter:
the mean metallicity for GCs increases as the GCs become young.
It is also noted that there are both MP and MR GCs in the old age, while there are only
MR GCs in the young age.
In Table \ref{tab-agemetal}, 
we derived the mean values of metallicity and age for the MR and MP GCs in the young group (age $<10$ Gyr) and old group (age $>10$ Gyr) in the combined sample of gEs and the MW. 
The number ratio of the young group and the old group is 1:3 for the MP GCs, and 1:1 for  the MR GCs.
In the case of the old group, the mean ages for the MR and MP GCs are about 13 Gyrs, and there is little
difference in the mean ages  between the MR and MP GCs. It is noted that the mean age for the MR GCs in the MW  is about one Gyr younger than that for the MP GCs.
In the case of the young group, the mean age for the MR GCs is about 5 Gyr, which is about 2 Gyr younger than that for the MP GCs.

\citet{puz05b} reported an existence of the age-metallicity relation 
 based on the high quality spectra for 17 GCs in seven early-type galaxies.
%
The age-metallicity relation shown for gE GCs as well as for MW GCs is also seen in the studies
 of the GCs in the Large Magellanic Cloud
 using the high-resolution spectroscopy \citep{col11}, 
 and of the GCs in six nearby galaxies 
 using the medium-resolution spectroscopy \citep{sha10}.
Therefore, it would be very interesting to examine 
 whether there is a global age-metallicity relation of GCs 
from dwarf galaxies to massive galaxies.  
 
Recently, \citet{mur10} presented a semi-analytic model for the origin of 
  the metallicity distribution of GCs
  using the galaxy merging 
    history from the cosmological simulations 
  coupled with observed scaling relations 
  for the amount and metallicity of cold gas available for the star formation.
  They predicted that
  early mergers of small galaxies create exclusively MP GCs,
  while subsequent mergers with more massive galaxies create both MP and MR GCs. 
   Thus they  showed the age-metallicity relation,
   which is that MR GCs are several Gyr younger  than MP GCs.  
   They also expected 
   that the mergers with more massive galaxies like the case of gEs
   would produce comparable numbers of MR and MP GCs simultaneously.
Our results, that the MR GCs in gEs are on average $\sim$ 3 Gyr younger than the MP GCs
  and the fraction of young MR GCs is non-negligible, 
   seem to be consistent with the model prediction by \citet{mur10}.
However, our result that there are comparable amounts of old MR GCs in gEs
  is not consistent with their model prediction 
  that exclusively old MP GCs formed in the early stage.

\citet{yoo11b} suggested nonlinear color-metallicity relations 
  between the colors of GCs and their metallicities.
  The shape of the metallicity distribution function (MDF) for GCs, 
   characterized by a sharp peak with a metal-poor tail, 
   indicates a continuous chemical enrichment with a short timescale less than 1 Gyr. 
They further suggested a possible systematic age difference among GC systems, 
  in that the GC systems in more luminous galaxies are older. 
Their result related to the MDF shape is not consistent with our result that
  the old GCs show a 
  bimodal metallicity distribution.
In addition, their result about the systematic  age difference among GC systems is not 
consistent with our result in the sense that 
 the mean ages of GCs in gEs do not show any systematic dependence on the luminosity of gEs
 as shown in Figure \ref{fig-gEsmagage}. 

Previously, several formation models of the GC systems in gEs have been suggested 
 \citep{pee69, ash92, har95, for97, cot98} and a summary of model descriptions and predictions can be found in several literature 
  \citep{rho01, lee03, ric04, wes04, bro06, hwa08, lee10a}. 
Considering  the observational results on the kinematics and photometry of GCs in gEs and  predictions of several models available in the literature, 
\citet{lee10a} presented a mixture (bibimbap) scenario describing the formation of GCs in gEs.
However, they did not use 
  the information on chemical properties and ages of GCs in gEs 
  when they proposed their scenario. 
  To be short, the scenario is as follows.
(1) MP GCs are formed mostly in low-mass dwarf galaxies very early, and preferentially in dwarf galaxies located in the high density environment like galaxy clusters.
 These are the first generation of GCs in the universe. MP GCs should be also formed in massive galaxies as well, but the number of these massive galaxies is much smaller than that of 
 the dwarf galaxies.
(2) MR GCs are formed together with stars in massive galaxies or 
 dissipational merging galaxies later than MP GCs, but not much later than MP GCs.  The chemical enrichment of the galaxies is rapid after the formation of MP GCs and the difference in the formation epoch of MP GCs and MR GCs should be small.  
(3) Massive galaxies grow becoming gEs via dissipationless or dissipational  merging of galaxies of various types and via accretion of many dwarf galaxies. New MR GCs will be formed during the dissipational merging, but the fraction of dissipational merging at this stage should be minor.
A significant fraction of MP GCs in gEs we see today are from dissipationless merging or accretion.

If we consider our results on ages and chemical properties in the view of this mixture scenario,
  we can conclude the followings.
  First, the point in the mixture scenario of \citet{lee10a} 
      that MP GCs in massive galaxies are formed early 
      is consistent   with our finding that most MP GCs in gEs are older than 10 Gyr.
 Second, the small difference in the primary formation epochs of MP GCs and MR GCs in \citet{lee10a} 
      is consistent  with  our results  
     that there is little difference in the mean ages of the MP and MR sub-populations of old GCs ($>10$ Gyr). 
      Note that there is $\sim1$ Gyr difference in the case of the MW GCs that have more reliable age estimates
      ($12.9\pm0.3$ Gyr and $11.8\pm0.2$ Gyr for the MP and MR GCs, respectively).

Important findings in this study are that gEs have a significant number of young GCs and that
they are mostly MR GCs. We also found that there are a smaller number of MP GCs with young ages
($<10$ Gyr).           
The MR young GCs are probably formed during the dissipational merging of late-type galaxies with
their host elliptical galaxies, while the MP young GCs are formed in dwarf galaxies and are accreted to their
host galaxies.  
It is expected that the fraction of new MR GCs from the dissipational merging events
depends on the environment such as the gas content of 
the merging companions and the local density where the merging happens
 \citep{par09,hwa10}.
Therefore, the fraction of young GCs would not be the same for all galaxies.
In the case of NGC 4636,
    a considerable fraction  of young GCs with ages less than 5 Gyr might be 
    the result of recent merging with gaseous late-type galaxies.
  It would be worth   investigating the environmental dependence of the GC systems in gEs once
  the number of gEs with spectroscopically measured metallicity and age 
  for their GCs increases  in the future.


\section{Summary}

We presented the measurements of the metallicity, age, and \afe
  of GCs in NGC 4636 derived from the Subaru spectroscopic data.
For comparison we have also re-derived the metallicity, age, and \afe  of GCs  for six other gEs (M87, M49, M60, NGC 5128, NGC 1399, and NGC 1407) from the line indices in the literature using the same methods as used
for NGC 4636 GCs.
We made a combined sample of GCs in gEs by combining all the GC data in gEs including NGC 4636, 
and used it to investigate various properties of GCs in gEs. Final results
for the combined sample as well as individual galaxies are summarized in Tables 7 -- 11. 
Our primary results are summarized below.
 
\begin{enumerate}
  
\item We measured the metallicities of 59 GCs in NGC 4636 with the BH method, and
	the metallicity, age, and \afe values of 33 GCs with the grid method and $\chi^2$ minimization method.
	
\item The metallicities of the NGC 4636 GCs show marginally a bimodal distribution 
with two peaks at [Fe/H] = 
$-1.23(\sigma=0.32)$ and $-0.35(\sigma=0.19)$ from the BH method,
$-1.14(\sigma=0.22)$ and $-0.41(\sigma=0.21)$ from the grid method, and
$-1.22(\sigma=0.25)$ and $-0.44(\sigma=0.16)$ from the $\chi^2$ minimization method.
The mean values of metallicity are 
	[Fe/H] $=-0.90\pm0.12$ from the BH method, 
$-0.70\pm0.13$ from the grid method, and
$-0.74\pm0.11$ from the $\chi^2$ minimization method.
	
\item 
	The ages of the GCs in NGC 4636 show a large range with a non-negligible number of young GCs.
	The mean age of the GCs is derived to be $7.9 \pm 0.8$ Gyr from the grid method and
$8.7 \pm 2.0$ Gyr from the $\chi^2$ minimization method. 
The number ratio for young ($<5$ Gyr), intermediate ($5-10$ Gyr), and old ($>10$ Gyr) GCs is 
27:39:34 from the grid method and 27:30:43 from the $\chi^2$ minimization method. 

\item 
	The \afe distribution of the NGC 4636 GCs is broad 
	with the mean of 
	\afe $=0.20\pm0.11$ from the grid method and of \afe $=0.07\pm0.05$ from the $\chi^2$ minimization method, smaller than the value for the MW GCs.

\item The metallicity, age, and alpha-element of the GCs in NGC 4636
  do not show any significant radial variation.
  NGC 4636 GCs show a hint of the age-metallicity relation.
%
	
	
\item 
The GCs in the combined sample of gEs show a bimodal metallicity distribution: peaks at 
$\textrm{[Fe/H]}=-1.21 (\sigma=0.32)$, and $-0.42 (\sigma=0.27)$ from the grid method, and
$\textrm{[Fe/H]}=-1.25 (\sigma=0.32)$, and $-0.42 (\sigma=0.25)$ from the $\chi^2$ minimization method. 
The MP GCs are on average $\sim 3$ Gyr older than the MR GCs. 
They show a large range in age from 2 to 15 Gyr, including a significant number of young GCs, while all MW GCs are older than 7 Gyr.
The MR GCs show a broader age distribution than the MP GCs, showing that young GCs are mostly metal-rich.  
  The mean \afe value of the combined gE GCs is smaller than that of the MW GCs.
  
\item 
The number ratio of the young group and the old group is 1:3 for the MP GCs, and 1:1 for  the MR GCs.
In the case of the old group, the mean ages for the MR and MP GCs are about 13 Gyrs, and there is little
difference in the mean ages  between the MR and MP GCs. It is noted that the mean age for the MR GCs in the MW is about one Gyr younger than that for the MP GCs.
In the case of the young group, the mean age for the MR GCs is about 5 Gyr, which is about 2 Gyr younger than that for the MP GCs.
   
\item The results on metallicity, age, and \afe of GCs in gEs 
as well as those on photometry and kinematics of GCs 
are consistent with the mixture scenario in \citet{lee10a}.
  
\end{enumerate}


\acknowledgments 
The authors like to thank the referee for his/her useful comments which improved significantly
the original manuscript. 
The authors are grateful to the staff of the SUBARU Telescope
 for their kind help during the observation.
This is supported in part
by the Mid-career Researcher Program through an NRF grant funded by the MEST (No.2010-0013875).
H.S.H. acknowledges the support of the Smithsonian Institution.





\begin{deluxetable}{cccr}
\tablewidth{0pc} 
\tablecaption{Zero Point Offsets for Lick Index calibration 
\label{tab-translick}}
\tablehead{
\colhead{index} & \colhead{unit} & \colhead{offset\tablenotemark{a}} & \colhead{rms} }
\startdata
     CN1  & mag &       --0.021  & 0.004 \\
     CN2  & mag &       --0.007  & 0.018 \\
  Ca4227  & \AA &        0.021  & 0.367 \\
   G4300  & \AA &        0.414  & 0.249 \\
  Fe4383  & \AA &        0.515  & 0.259 \\
  Ca4455  & \AA &        0.408  & 0.291 \\
  Fe4531  & \AA &        0.842  & 0.780 \\
  C24668  & \AA &       --1.880  & 1.036 \\
   H$\beta$  & \AA &        0.184  & 0.140 \\
  Fe5015  & \AA &        0.616  & 0.267 \\
     Mg1  & mag &       --0.017  & 0.016 \\
     Mg2  & mag &       --0.015  & 0.018 \\
     Mgb  & \AA &        0.165  & 0.187 \\
  Fe5270  & \AA &        0.443  & 0.306 \\
  Fe5335  & \AA &       --0.103  & 0.162 \\
  Fe5406  & \AA &        0.271  & 0.147 \\
  Fe5709  & \AA &        0.148  & 0.484 \\
  Fe5782  & \AA &        0.036  & 0.156 \\
     NaD  & \AA &        0.494  & 0.578 \\
 H$\delta_A$  & \AA &       --0.367  & 0.458 \\
 H$\gamma_A$  & \AA &       --0.761  & 0.645 \\
 H$\delta_F$  & \AA &        0.129  & 0.242 \\
 H$\gamma_F$  & \AA &        0.083  & 0.269 \\
\enddata
\tablenotetext{a~}{$Index$(Lick)=$Index$(Subaru)+$constant$.} 
\end{deluxetable}
\clearpage

\begin{deluxetable}{crrrrrrrrrrrrrrrrrrrrrrr} 
\tabletypesize{\tiny} 
\tablewidth{0pc} 
\tablecaption{Lick Line Indices (set 1) \label{tab-lickindex1}}
\tablehead{ 
\colhead{ID\tablenotemark{a}} & 
\colhead{CN1} & \colhead{CN2} & \colhead{Ca4227} & \colhead{G4300} & \colhead{Fe4383} &
\colhead{Ca4455} & \colhead{Fe4531} & \colhead{C24668} & \colhead{H$\beta$} & \colhead{Fe5015} &
\colhead{Mg1} & \colhead{Mg2} \\ 
\colhead{ } &
\colhead{(mag)} & \colhead{(mag)} & \colhead{(\AA} & \colhead{(\AA)} & \colhead{(\AA)} &
\colhead{(\AA)}  & \colhead{(\AA)} & \colhead{(\AA)} & \colhead{(\AA)} & \colhead{(\AA)} &
\colhead{(mag)} & \colhead{(mag)} }
\startdata
    157 &   0.219 &   0.242 &   0.747 &   4.612 &   6.693 &  --0.054 &   4.808 &  --2.320 &   1.865 &   4.851 &   0.093 &   0.219 \\ 
    182 &  --0.070 &  --0.030 &   1.912 &   0.824 &  --1.213 &   0.581 &   0.719 &  --3.783 &   1.935 &  --0.530 &   0.001 &   0.073 \\ 
    244 &   0.290 &   0.322 &  --3.348 &   6.267 &   5.615 &   3.151 &   1.416 &  --0.785 &   2.230 &   3.279 &   0.056 &   0.233 \\ 
    ...
\enddata
\tablenotetext{a~}{From \citet{par10,par12}}
\tablecomments{This table is available in its entirety in amachine-readable form in the online journal.A portion is shown here for guidance regarding its form and content.}
\end{deluxetable}

\begin{deluxetable}{crrrrrrrrrrrrrrrrrrrrrrr} 
\tabletypesize{\tiny} 
\tablewidth{0pc} 
\tablecaption{Lick Line Indices (set 2) \label{tab-lickindex2}}
\tablehead{ 
\colhead{ID\tablenotemark{a}} & 
\colhead{Mgb} & \colhead{Fe5270} & \colhead{Fe5335} &
\colhead{Fe5406} & \colhead{Fe5709} & \colhead{Fe5782} & \colhead{NaD} & \colhead{H$\delta_A$} &
\colhead{H$\gamma_A$} & \colhead{H$\delta_F$} & \colhead{H$\gamma_F$} \\ 
\colhead{ } &
\colhead{(\AA)} & \colhead{(\AA)} & \colhead{(\AA)} &
\colhead{(\AA)} & \colhead{(\AA)} & \colhead{(\AA)} & \colhead{(\AA)} & \colhead{(\AA)} &
\colhead{(\AA)} & \colhead{(\AA)} & \colhead{(\AA)} }
\startdata
    157 &   3.837 &   2.943 &   2.952 &   2.993 &   1.325 &   0.556 &   1.426 &  --3.797 &  --5.303 &  --0.301 &  --0.951 \\
    182 &   0.322 &   1.199 &   2.723 &  -0.436 &   0.916 &   0.360 &   5.777 &  --0.220 &   3.506 &  --0.837 &   2.851 \\ 
    244 &   5.220 &   0.723 &   2.289 &   1.335 &   1.666 &  --0.001 &   1.963 &  --3.485 & --17.252 &  --1.465 &  --7.051 \\ 
    ...
\enddata
\tablenotetext{a~}{From \citet{par10,par12}}
\tablecomments{This table is available in its entirety in amachine-readable form in the online journal.A portion is shown here for guidance regarding its form and content.}
\end{deluxetable}

\begin{deluxetable}{crrrrrrrrrrrrrrrrrrrrrrr} 
\tabletypesize{\tiny} 
\tablewidth{0pc} 
\tablecaption{Lick Index Errors  (set 1) \label{tab-lickerr1}}
\tablehead{ 
\colhead{ID\tablenotemark{a}} & 
\colhead{CN1} & \colhead{CN2} & \colhead{Ca4227} & \colhead{G4300} & \colhead{Fe4383} &
\colhead{Ca4455} & \colhead{Fe4531} & \colhead{C24668} & \colhead{H$\beta$} & \colhead{Fe5015} &
\colhead{Mg1} & \colhead{Mg2} \\ 
\colhead{ } &
\colhead{(mag)} & \colhead{(mag)} & \colhead{(\AA} & \colhead{(\AA)} & \colhead{(\AA)} &
\colhead{(\AA)}  & \colhead{(\AA)} & \colhead{(\AA)} & \colhead{(\AA)} & \colhead{(\AA)} &
\colhead{(mag)} & \colhead{(mag)} }
\startdata
    157 &   0.024 &   0.031 &   0.502 &   0.931 &   1.217 &   0.553 &   0.898 &   1.341 &   0.447 &   1.013 &   0.014 &   0.016 \\ 
    182 &   0.019 &   0.025 &   0.404 &   0.790 &   1.188 &   0.502 &   0.916 &   1.367 &   0.439 &   1.019 &   0.014 &   0.016 \\
    244 &   0.028 &   0.036 &   0.614 &   1.168 &   1.513 &   0.633 &   1.082 &   1.574 &   0.502 &   1.143 &   0.016 &   0.012 \\ 
    ...
\enddata
\tablenotetext{a~}{From \citet{par10,par12}}
\tablecomments{This table is available in its entirety in amachine-readable form in the online journal.A portion is shown here for guidance regarding its form and content.}
\end{deluxetable}

\begin{deluxetable}{crrrrrrrrrrrrrrrrrrrrrrr} 
\tabletypesize{\tiny} 
\tablewidth{0pc} 
\tablecaption{Lick Index Errors (set 2) \label{tab-lickerr2}}
\tablehead{ 
\colhead{ID\tablenotemark{a}} & 
\colhead{Mgb} & \colhead{Fe5270} & \colhead{Fe5335} &
\colhead{Fe5406} & \colhead{Fe5709} & \colhead{Fe5782} & \colhead{NaD} & \colhead{H$\delta_A$} &
\colhead{H$\gamma_A$} & \colhead{H$\delta_F$} & \colhead{H$\gamma_F$} \\ 
\colhead{ } &
\colhead{(\AA)} & \colhead{(\AA)} & \colhead{(\AA)} &
\colhead{(\AA)} & \colhead{(\AA)} & \colhead{(\AA)} & \colhead{(\AA)} & \colhead{(\AA)} &
\colhead{(\AA)} & \colhead{(\AA)} & \colhead{(\AA)} }
\startdata
    157 &   0.457 &   0.461 &   0.581 &   0.394 &   0.275 &   0.266 &   0.424 &   1.013 &   0.756 &   0.647 &   0.438 \\ 
    182 &   0.435 &   0.454 &   0.567 &   0.409 &   0.275 &   0.278 &   0.423 &   0.848 &   0.655 &   0.562 &   0.397 \\
    244 &   0.516 &   0.525 &   0.678 &   0.444 &   0.308 &   0.289 &   0.437 &   1.148 &   0.918 &   0.751 &   0.508 \\ 
    ...
\enddata
\tablenotetext{a~}{From \citet{par10,par12}}
\tablecomments{This table is available in its entirety in amachine-readable form in the online journal.A portion is shown here for guidance regarding its form and content.}
\end{deluxetable}

\begin{deluxetable}{ccccccccc}
\tabletypesize{\scriptsize} 
\tablewidth{0pc} 
\tablecaption{Metallicity, Age, and [$\alpha$/Fe] of NGC 4636 GCs\label{tab-catalog}}
\tablehead{ 
\colhead{ID\tablenotemark{a}} & 
\colhead{[Z/H]$_{grid}^{[MgFe]^\prime}$} & 
\colhead{[Z/H]$_{grid}^{Mg2}$} &
\colhead{Age$_{grid}$} &
\colhead{[$\alpha$/Fe]$_{grid}$} &
\colhead{[Z/H]$_{\chi^2}$} & 
\colhead{Age$_{\chi^2}$} &
\colhead{[$\alpha$/Fe]$_{\chi^2}$} &
\colhead{[Fe/H]$_{BH}$} \\
\colhead{} & 
\colhead{(dex)} & 
\colhead{(dex)} &
\colhead{(Gyr)} &
\colhead{(dex)} &
\colhead{(dex)} &
\colhead{(Gyr)} &
\colhead{(dex)} &
\colhead{(dex)}
 }
\startdata 
      157 & $  0.27 \pm   0.28$ & $  0.11 \pm   0.05$ &$    4.7 \pm    3.9 $& $ -0.08 \pm   0.10 $& $  0.11 \pm   0.09 $& $   7.0 \pm    2.6$ & $ -0.20 \pm   0.12 $& $ -0.13 \pm   0.16 $\\ 
    182 & $ -1.70 \pm   0.39$ & $ -1.04 \pm   0.10$ &$   14.0 \pm    1.5 $& $ -0.30 \pm   0.00 $& $ -1.46 \pm   0.23 $& $  12.0 \pm    3.6$ & $ -0.30 \pm   0.25 $& $ -1.79 \pm   0.16 $\\ 
    244 & $ -0.21 \pm   0.22$ & $ -0.12 \pm   0.05$ &$    5.4 \pm    4.0 $& $  0.49 \pm   0.02 $& $  0.23 \pm   0.09 $& $   2.2 \pm    0.8$ & $  0.26 \pm   0.21 $& $ -0.32 \pm   0.48 $\\ 
    ...
\enddata
\tablenotetext{a~}{From \citet{par10,par12}}
\tablecomments{This table is available in its entirety in amachine-readable form in the online journal.A portion is shown here for guidance regarding its form and content.}
\end{deluxetable}

\begin{deluxetable}{cccccc}
\tabletypesize{\scriptsize} 
\tablewidth{0pc} 
\tablecaption{KMM Test Results for 
the Metallicities of GCs in gEs\tablenotemark{a}\label{tab-metalpeak}}
\tablehead{ 
\colhead{galaxy} & 
\colhead{N} & \colhead{[Fe/H]$_{MP}$ $\pm$ $\sigma$} &
\colhead{N} & \colhead{[Fe/H]$_{MR}$ $\pm$ $\sigma$} &
\colhead{confidence level} \\
\colhead{} &
\colhead{} & \colhead{(dex)} &
\colhead{} & \colhead{(dex)} &
\colhead{(\%)} 
}
\startdata
\multicolumn{6}{c}{\underbar{grid method}}\\ 
   NGC 4636 &   10 & $ -1.14\pm 0.22 $ &   15 & $ -0.41\pm 0.21 $ &  71.4 \\
   NGC 5128 &   42 & $ -1.14\pm 0.32 $ &   14 & $ -0.34\pm 0.12 $ &  97.7 \\
     M60 &    5 & $ -1.27\pm 0.62 $ &   33 & $ -0.56\pm 0.36 $ &  71.9 \\
   NGC 1407 &    7 & $ -1.14\pm 0.23 $ &   12 & $ -0.38\pm 0.18 $ &  76.2 \\
   NGC 1399 &    3 & $ -1.63\pm 0.06 $ &    7 & $ -0.25\pm 0.51 $ &  94.6 \\
     M87 &   66 & $ -1.24\pm 0.32 $ &   63 & $ -0.45\pm 0.27 $ &  97.5 \\
     M49 &   28 & $ -0.82\pm 0.65 $ &   15 & $  0.37\pm 0.10 $ &  99.9 \\     
     gEs &  155 & $ -1.21\pm 0.32 $ &  149 & $ -0.42\pm 0.27 $ &  99.9 \\
     gEs ($\le$ 10 Gyr) &   22 & $ -1.41\pm 0.14 $ &   94 & $ -0.57\pm 0.37 $ &  98.4 \\
     gEs ($>  $ 10 Gyr) &  111 & $ -1.22\pm 0.34 $ &   77 & $ -0.42\pm 0.25 $ &  99.7 \\
\multicolumn{6}{c}{}\\
\multicolumn{6}{c}{\underbar{$\chi^2$ minimization method}} \\    
   NGC 4636 &   14 & $ -1.22\pm 0.25 $ &   14 & $ -0.44\pm 0.16 $ &  95.2 \\
   NGC 5128 &   43 & $ -1.40\pm 0.47 $ &   19 & $ -0.43\pm 0.15 $ &  98.2 \\
     M60 &   20 & $ -1.09\pm 0.31 $ &   15 & $ -0.37\pm 0.11 $ &  98.8 \\
   NGC 1407 &    8 & $ -1.05\pm 0.33 $ &   11 & $ -0.39\pm 0.11 $ &  92.8 \\
   NGC 1399 &    4 & $ -1.68\pm 0.24 $ &    6 & $ -0.37\pm 0.21 $ &  94.5 \\
     M87 &   40 & $ -1.41\pm 0.14 $ &   90 & $ -0.67\pm 0.36 $ &  99.9 \\
     M49 &   26 & $ -1.15\pm 0.71 $ &   20 & $  0.19\pm 0.25 $ &  99.2 \\
     gEs &  167 & $ -1.25\pm 0.32 $ &  148 & $ -0.42\pm 0.25 $ &  99.9 \\
     gEs ($\le$ 10 Gyr) &   63 & $ -1.11\pm 0.43 $ &   64 & $ -0.32\pm 0.27 $ &  98.5 \\ 
     gEs ($>  $ 10 Gyr) &  114 & $ -1.25\pm 0.29 $ &   74 & $ -0.44\pm 0.18 $ &  99.9 \\  
\multicolumn{6}{c}{}\\
      MW\tablenotemark{b} &  111 & $ -1.52\pm 0.39 $ &   41 & $ -0.52\pm 0.23 $ & 99.8 \\     
\enddata
\tablenotetext{a~}{Peak values, widths ($\sigma$), and confidence levels are derived from the KMM test.}
\tablenotetext{b~}{It is from \citet{har96}.}
\end{deluxetable}

\begin{deluxetable}{ccccccccccc}
\rotate
\tabletypesize{\scriptsize} 
\tablewidth{0pc} 
\tablecaption{Mean Metallicities of GCs in gEs\tablenotemark{a}\label{tab-metaldist}}
\tablehead{ 
\colhead{} & 
\colhead{} &
\colhead{total} &
\colhead{} &
\colhead{age $\le$ 10 Gyr} &
\colhead{} &
\colhead{age $>$ 10 Gyr} &
\colhead{} &
\colhead{[$\alpha$/Fe] $\le$ 0.2} &
\colhead{} &
\colhead{[$\alpha$/Fe] $>$ 0.2} \\ 
\colhead{galaxy} & 
\colhead{N} &
\colhead{$<$[Fe/H]$>$ $\pm$ $<\sigma>$} &
\colhead{N} &
\colhead{$<$[Fe/H]$>$ $\pm$ $<\sigma>$} &
\colhead{N} &
\colhead{$<$[Fe/H]$>$ $\pm$ $<\sigma>$} &
\colhead{N} &
\colhead{$<$[Fe/H]$>$ $\pm$ $<\sigma>$} &
\colhead{N} &
\colhead{$<$[Fe/H]$>$ $\pm$ $<\sigma>$} \\ 
\colhead{} &
\colhead{} &
\colhead{(dex)} &
\colhead{} &
\colhead{(dex)} &
\colhead{} &
\colhead{(dex)} &
\colhead{} &
\colhead{(dex)} &
\colhead{} &
\colhead{(dex)}
}
\startdata
\multicolumn{11}{c}{\underbar{grid method}}\\ 
   NGC 4636 &  33&$ -0.70\pm  0.13$&  22&$ -0.49\pm  0.14$&  11&$ -1.20\pm  0.25$&  17&$ -0.50\pm  0.18$&  16&$ -0.86\pm  0.21$ \\
   NGC 5128 &  68&$ -1.10\pm  0.08$&  23&$ -0.81\pm  0.11$&  45&$ -1.27\pm  0.11$&  44&$ -1.07\pm  0.11$&  24&$ -1.14\pm  0.13$ \\
     M60 &  38&$ -0.67\pm  0.08$&  10&$ -0.93\pm  0.20$&  28&$ -0.59\pm  0.09$&  24&$ -0.70\pm  0.09$&  14&$ -0.58\pm  0.18$ \\
   NGC 1407 &  20&$ -0.68\pm  0.13$&   4&$ -0.48\pm  0.20$&  16&$ -0.74\pm  0.15$&   1&$ -2.39\pm  0.16$&  19&$ -0.65\pm  0.13$ \\
   NGC 1399 &  10&$ -0.66\pm  0.27$&   7&$ -0.33\pm  0.47$&   3&$ -1.60\pm  0.43$&   7&$ -0.93\pm  0.49$&   3&$  0.32\pm  0.62$ \\
     M87 & 146&$ -0.83\pm  0.06$&  56&$ -0.57\pm  0.08$&  90&$ -1.00\pm  0.07$& ...&$  ...          $& ...&$           ... $ \\
     M49 &  43&$ -0.42\pm  0.14$&  20&$ -0.10\pm  0.34$&  23&$ -0.68\pm  0.20$&  30&$ -0.48\pm  0.16$&  13&$ -0.22\pm  0.30$ \\
      gEs & 358&$ -0.80\pm  0.04$& 142&$ -0.55\pm  0.05$& 216&$ -0.96\pm  0.05$& 123&$ -0.78\pm  0.08$&  89&$ -0.76\pm  0.08$ \\
\multicolumn{11}{c}{}\\
\multicolumn{11}{c}{\underbar{$\chi^2$ minimization method}} \\ 
   NGC 4636 &  33&$ -0.74\pm  0.11$&  19&$ -0.59\pm  0.16$&  14&$ -0.94\pm  0.14$&  20&$ -0.92\pm  0.13$&  13&$ -0.40\pm  0.11$ \\
   NGC 5128 &  62&$ -1.15\pm  0.08$&  21&$ -1.11\pm  0.14$&  41&$ -1.17\pm  0.09$&  46&$ -1.20\pm  0.10$&  16&$ -1.01\pm  0.14$ \\
     M60 &  38&$ -0.76\pm  0.09$&  11&$ -0.58\pm  0.26$&  27&$ -0.82\pm  0.11$&  16&$ -0.75\pm  0.15$&  22&$ -0.76\pm  0.13$ \\
   NGC 1407 &  20&$ -0.62\pm  0.13$&   6&$ -0.52\pm  0.16$&  14&$ -0.69\pm  0.17$&   6&$ -0.91\pm  0.19$&  14&$ -0.44\pm  0.12$ \\
   NGC 1399 &  10&$ -0.83\pm  0.40$&   7&$ -0.51\pm  0.36$&   3&$ -1.16\pm  0.79$&   5&$ -1.48\pm  0.78$&   5&$ -0.56\pm  0.20$ \\
     M87 & 150&$ -0.90\pm  0.05$&  52&$ -0.63\pm  0.08$&  98&$ -1.08\pm  0.06$&  77&$ -1.02\pm  0.07$&  73&$ -0.77\pm  0.08$ \\
     M49 &  46&$ -0.58\pm  0.16$&  24&$ -0.44\pm  0.18$&  22&$ -0.77\pm  0.25$&  23&$ -0.69\pm  0.19$&  23&$ -0.38\pm  0.29$ \\
      gEs & 359&$ -0.86\pm  0.04$& 140&$ -0.65\pm  0.06$& 219&$ -1.00\pm  0.04$& 193&$ -1.00\pm  0.05$& 166&$ -0.70\pm  0.05$ \\
\multicolumn{11}{c}{}\\
      MW & 152&$ -1.28\pm  0.05$&   3&$ -0.61\pm  0.27$&  61&$ -1.44\pm  0.08$&   8&$ -1.33\pm  0.30$&  49&$ -1.40\pm  0.10$ \\
\enddata
\tablenotetext{a~}{The mean values are determined with the biweight location of \citet{bee90}.
The uncertainties, $<\sigma>$, indicate the 68\% confidence levels obtained with a bootstrap method.}
\end{deluxetable}

\begin{deluxetable}{ccccccccccc}
\rotate
\tabletypesize{\scriptsize} 
\tablewidth{0pc} 
\tablecaption{Mean Ages of GCs in gEs\tablenotemark{a}\label{tab-agedist}}
\tablehead{ 
\colhead{} & 
\colhead{} &
\colhead{total} &
\colhead{} &
\colhead{[Fe/H] $\le$ --0.9} &
\colhead{} &
\colhead{[Fe/H] $>$ --0.9} &
\colhead{} &
\colhead{[$\alpha$/Fe] $\le$ 0.2} &
\colhead{} &
\colhead{[$\alpha$/Fe] $>$ 0.2} \\ 
\colhead{galaxy} & 
\colhead{N} &
\colhead{$<$age$>$ $\pm$ $<\sigma>$} &
\colhead{N} &
\colhead{$<$age$>$ $\pm$ $<\sigma>$} &
\colhead{N} &
\colhead{$<$age$>$ $\pm$ $<\sigma>$} &
\colhead{N} &
\colhead{$<$age$>$ $\pm$ $<\sigma>$} &
\colhead{N} &
\colhead{$<$age$>$ $\pm$ $<\sigma>$} \\ 
\colhead{} &
\colhead{} &
\colhead{(Gyr)} &
\colhead{} &
\colhead{(Gyr)} &
 \colhead{} &
\colhead{(Gyr)} &
\colhead{} &
\colhead{(Gyr)} &
\colhead{} &
\colhead{(Gyr)}
}
\startdata  
\multicolumn{11}{c}{\underbar{grid method}}\\
   NGC 4636 &  33&$   7.9\pm   0.8$&  13&$  12.3\pm   2.1$&  20&$   6.4\pm   1.0$&  17&$   7.5\pm   1.3$&  16&$   8.2\pm   1.0$ \\
   NGC 5128 &  68&$  12.3\pm   0.5$&  42&$  12.4\pm   0.4$&  26&$  10.1\pm   1.8$&  44&$  12.3\pm   0.6$&  24&$  12.3\pm   0.5$ \\
     M60 &  38&$  13.0\pm   0.0$&  12&$  12.9\pm   1.6$&  26&$  13.0\pm   0.3$&  24&$  13.0\pm   0.3$&  14&$  12.6\pm   0.6$ \\
   NGC 1407 &  20&$  12.6\pm   0.3$&   8&$  12.0\pm   0.2$&  12&$  13.0\pm   0.1$&   1&$  13.0\pm   0.3$&  19&$  12.5\pm   0.4$ \\
   NGC 1399 &  10&$   6.9\pm   1.5$&   3&$  12.5\pm   4.0$&   7&$   5.5\pm   1.6$&   7&$   7.6\pm   1.7$&   3&$   4.9\pm   4.8$ \\
     M87 & 146&$  12.4\pm   0.2$&  66&$  12.1\pm   0.2$&  80&$   8.8\pm   2.2$& ...&$  ...          $& ...&$   ...         $ \\
     M49 &  43&$   9.5\pm   2.1$&  13&$  12.1\pm   0.2$&  30&$   7.7\pm   1.2$&  30&$   8.0\pm   1.3$&  13&$  12.2\pm   0.6$ \\
      gEs & 358&$  12.3\pm   0.1$& 157&$  12.2\pm   0.1$& 201&$   9.2\pm   0.7$& 123&$  12.4\pm   1.5$&  89&$  12.2\pm   0.2$ \\
\multicolumn{11}{c}{}\\
\multicolumn{11}{c}{\underbar{$\chi^2$ minimization method}} \\ 
   NGC 4636 &  33&$   8.7\pm   2.0$&  14&$  11.9\pm   1.5$&  19&$   7.3\pm   1.2$&  20&$   9.0\pm   1.0$&  13&$   8.2\pm   3.4$ \\
   NGC 5128 &  62&$  12.1\pm   0.4$&  40&$  12.3\pm   0.3$&  22&$  11.9\pm   1.3$&  46&$  12.4\pm   0.5$&  16&$  11.5\pm   1.0$ \\
     M60 &  38&$  12.9\pm   0.3$&  16&$  12.6\pm   0.4$&  22&$  13.0\pm   0.6$&  16&$  12.7\pm   1.3$&  22&$  12.8\pm   0.2$ \\
   NGC 1407 &  20&$  12.8\pm   0.5$&   6&$  12.9\pm   0.3$&  14&$  12.6\pm   1.4$&   6&$  10.0\pm   2.1$&  14&$  13.0\pm   0.2$ \\
   NGC 1399 &  10&$   7.5\pm   1.8$&   4&$  11.5\pm   2.0$&   6&$   5.1\pm   1.7$&   5&$   9.5\pm   3.3$&   5&$   4.8\pm   2.6$ \\
     M87 & 150&$  12.3\pm   0.1$&  75&$  12.2\pm   0.1$&  75&$  11.7\pm   2.1$&  77&$  12.3\pm   0.1$&  73&$  12.4\pm   0.1$ \\
     M49 &  46&$   9.5\pm   1.5$&  18&$  11.7\pm   1.0$&  28&$   8.6\pm   1.2$&  23&$   9.0\pm   1.0$&  23&$  11.8\pm   1.5$ \\
      gEs & 359&$  12.2\pm   0.1$& 173&$  12.2\pm   0.1$& 186&$   9.7\pm   0.7$& 193&$  12.2\pm   0.1$& 166&$  12.3\pm   0.1$ \\
\multicolumn{11}{c}{}\\
      MW &  64&$  12.2\pm   0.1$&  48&$  12.9\pm   0.3$&  16&$  11.8\pm   0.2$&   6&$  10.7\pm   1.0$&  38&$  13.0\pm   0.2$ \\
\enddata
\tablenotetext{a~}{The mean values are determined with the biweight location of \citet{bee90}.
The uncertainties, $<\sigma>$, indicate the 68\% confidence levels obtained with a bootstrap method.}

\end{deluxetable}

\begin{deluxetable}{ccccccccccc}
\rotate
\tabletypesize{\scriptsize} 
\tablewidth{0pc} 
\tablecaption{Mean \afee s of GCs in gEs\tablenotemark{a}\label{tab-alphadist}}
\tablehead{ 
\colhead{} & 
\colhead{} &
\colhead{total} &
\colhead{} &
\colhead{[Fe/H] $\le$ --0.9} &
\colhead{} &
\colhead{[Fe/H] $>$ --0.9} &
\colhead{} &
\colhead{age $\le$ 10 Gyr} &
\colhead{} &
\colhead{age $>$ 10 Gyr} \\ 
\colhead{galaxy} & 
\colhead{N} &
\colhead{$<$\afe$>$ $\pm$ $<\sigma>$} &
\colhead{N} &
\colhead{$<$\afe$>$ $\pm$ $<\sigma>$} &
\colhead{N} &
\colhead{$<$\afe$>$ $\pm$ $<\sigma>$} &
\colhead{N} &
\colhead{$<$\afe$>$ $\pm$ $<\sigma>$} &
\colhead{N} &
\colhead{$<$\afe$>$ $\pm$ $<\sigma>$} \\ 
\colhead{} &
\colhead{} &
\colhead{(dex)} &
\colhead{} &
\colhead{(dex)} &
 \colhead{} &
\colhead{(dex)} &
\colhead{} &
\colhead{(dex)} &
\colhead{} &
\colhead{(dex)}
}
\startdata
\multicolumn{11}{c}{\underbar{grid method}}\\
   NGC 4636 &  33&$  0.20\pm  0.11$&  13&$  0.48\pm  0.15$&  20&$  0.17\pm  0.06$&  22&$  0.22\pm  0.14$&  11&$  0.17\pm  0.20$ \\
   NGC 5128 &  68&$  0.13\pm  0.03$&  42&$  0.14\pm  0.04$&  26&$  0.12\pm  0.05$&  23&$  0.14\pm  0.05$&  45&$  0.12\pm  0.04$ \\
     M60 &  38&$  0.06\pm  0.04$&  12&$  0.01\pm  0.22$&  26&$  0.08\pm  0.05$&  10&$ -0.08\pm  0.15$&  28&$  0.09\pm  0.05$ \\
   NGC 1407 &  20&$  0.42\pm  0.05$&   8&$  0.44\pm  0.05$&  12&$  0.40\pm  0.06$&   4&$  0.36\pm  0.06$&  16&$  0.46\pm  0.05$ \\
   NGC 1399 &  10&$  0.07\pm  0.09$&   3&$ -0.30\pm  0.19$&   7&$  0.21\pm  0.08$&   7&$  0.09\pm  0.10$&   3&$  0.09\pm  0.39$ \\
     M87 & ...&$  ...          $& ...&$  ...          $& ...&$           ... $& ...&$  ...          $& ...&$ ...           $ \\
     M49 &  43&$  0.09\pm  0.04$&  13&$ -0.04\pm  0.19$&  30&$  0.13\pm  0.05$&  20&$  0.02\pm  0.05$&  23&$  0.15\pm  0.07$ \\
      gEs & 212&$  0.14\pm  0.02$&  91&$  0.14\pm  0.03$& 121&$  0.15\pm  0.02$&  86&$  0.12\pm  0.03$& 126&$  0.16\pm  0.02$ \\
\multicolumn{11}{c}{}\\
\multicolumn{11}{c}{\underbar{$\chi^2$ minimization method}} \\ 
   NGC 4636 &  33&$  0.07\pm  0.05$&  14&$ -0.10\pm  0.07$&  19&$  0.21\pm  0.07$&  19&$  0.05\pm  0.06$&  14&$  0.10\pm  0.11$ \\
   NGC 5128 &  62&$  0.06\pm  0.03$&  40&$  0.02\pm  0.04$&  22&$  0.12\pm  0.03$&  21&$  0.07\pm  0.06$&  41&$  0.05\pm  0.03$ \\
     M60 &  38&$  0.26\pm  0.05$&  16&$  0.36\pm  0.15$&  22&$  0.26\pm  0.03$&  11&$  0.24\pm  0.08$&  27&$  0.28\pm  0.05$ \\
   NGC 1407 &  20&$  0.31\pm  0.03$&   6&$  0.21\pm  0.13$&  14&$  0.33\pm  0.04$&   6&$  0.25\pm  0.07$&  14&$  0.33\pm  0.04$ \\
   NGC 1399 &  10&$  0.15\pm  0.09$&   4&$ -0.25\pm  0.15$&   6&$  0.32\pm  0.15$&   7&$  0.18\pm  0.19$&   3&$  0.17\pm  0.18$ \\
     M87 & 150&$  0.16\pm  0.02$&  75&$  0.05\pm  0.04$&  75&$  0.24\pm  0.02$&  52&$  0.21\pm  0.04$&  98&$  0.13\pm  0.03$ \\
     M49 &  46&$  0.19\pm  0.04$&  18&$  0.11\pm  0.14$&  28&$  0.24\pm  0.04$&  24&$  0.17\pm  0.06$&  22&$  0.23\pm  0.11$ \\
      gEs & 359&$  0.16\pm  0.01$& 173&$  0.07\pm  0.02$& 186&$  0.23\pm  0.01$& 140&$  0.16\pm  0.02$& 219&$  0.15\pm  0.02$ \\
\multicolumn{11}{c}{}\\
      MW &  57&$  0.36\pm  0.01$&  43&$  0.38\pm  0.01$&  14&$  0.29\pm  0.04$&   2&$ -0.02\pm  0.01$&  42&$  0.37\pm  0.01$ \\
\enddata
\tablenotetext{a~}{The mean values are determined with the biweight location of \citet{bee90}.
The uncertainties, $<\sigma>$, indicate the 68\% confidence levels obtained with a bootstrap method.}
\end{deluxetable}

\begin{deluxetable}{ccccccc}
\tablewidth{0pc} 
\tablecaption{Mean Age and Metallicity of GCs in gEs and the MW\tablenotemark{a}\label{tab-agemetal}}
\tablehead{ 
\colhead{} & 
\multicolumn{3}{c}{young GC (age $\le$ 10 Gyr)} & 
\multicolumn{3}{c}{old GC (age $>$ 10 Gyr)} \\
\colhead{} &
\colhead{N} &  \colhead{$<$[Fe/H]$>$} & \colhead{$<$age$>$} &
\colhead{N} &  \colhead{$<$[Fe/H]$>$} & \colhead{$<$age$>$} \\
\colhead{} &
\colhead{} &  \colhead{(dex)}  & \colhead{(Gyr)} & 
\colhead{} &  \colhead{(dex)}  & \colhead{(Gyr)}
}
\startdata
\multicolumn{7}{c}{\underbar{gEs (grid)}}\\
MP GC ([Fe/H] $\le$ --0.9) &  42&$ -1.26\pm  0.04$& $   7.0\pm   0.3$& 115&$ -1.41\pm  0.06$& $  12.5\pm   0.4$ \\ 
MR GC ([Fe/H] $>$ --0.9)   & 100&$ -0.27\pm  0.05$& $   4.8\pm   0.3$& 101&$ -0.45\pm  0.03$& $  13.0\pm   0.3$ \\ 
\multicolumn{7}{c}{}\\
\multicolumn{7}{c}{\underbar{gEs ($\chi^2$ minimization)}} \\ 
MP GC ([Fe/H] $\le$ --0.9) &  50&$ -1.33\pm  0.05$& $   7.2\pm   0.3$& 123&$ -1.39\pm  0.06$& $  12.5\pm   0.5$ \\ 
MR GC ([Fe/H] $>$ --0.9)   &  90&$ -0.32\pm  0.04$& $   5.1\pm   0.3$&  96&$ -0.48\pm  0.03$& $  13.0\pm   0.3$ \\ 
\multicolumn{7}{c}{}\\
\multicolumn{7}{c}{\underbar{MW}} \\ 
MP GC ([Fe/H] $\le$ --0.9) &  0& ...   & ...                        &  48&$ -1.64\pm  0.05$& $  12.9\pm   0.3$ \\ 
MR GC ([Fe/H] $>$ --0.9)   &  3&$ -0.61\pm  0.26$& $   7.0\pm   0.8$&  13&$ -0.60\pm  0.06$& $  11.8\pm   0.3$ \\ 
\enddata
\tablenotetext{a~}{The mean values are determined with the biweight location of \citet{bee90}.
The uncertainties indicate the 68\% confidence levels obtained with a bootstrap method.}
\end{deluxetable}

\clearpage

\clearpage

\clearpage

\clearpage

\clearpage

\clearpage

\clearpage

\clearpage

\clearpage

\clearpage

\clearpage

\clearpage

\clearpage

\clearpage

\clearpage

\end{document}